\documentclass[acmsmall]{acmart}
\usepackage{graphicx}

\usepackage{multirow}
\usepackage{ulem}

\AtBeginDocument{%
  \providecommand\BibTeX{{%
    \normalfont B\kern-0.5em{\scshape i\kern-0.25em b}\kern-0.8em\TeX}}}

\setcopyright{rightsretained}
\acmJournal{PACMHCI}
\acmYear{2024} \acmVolume{8} \acmNumber{CSCW1} \acmArticle{35} \acmMonth{4}\acmDOI{10.1145/3637312}
\usepackage{soul}
\usepackage{subcaption}
\usepackage{multirow}
\usepackage{csquotes}

\newcommand{\Comment}[1]{}

\usepackage{graphicx}
\usepackage{enumitem}
\setlist[description]{leftmargin=2\parindent,labelindent=2\parindent,rightmargin=2\parindent}

\setlist[itemize]{leftmargin=*}
\setlist[enumerate]{leftmargin=*}

\received{January 2023}
\received[revised]{July 2023}
\received[accepted]{November 2023}

\begin{document}

\title[Designing Information Artifacts that Bridge from Synchronous Meetings to Asynchronous Collaboration]{Meeting Bridges: Designing Information Artifacts that Bridge from Synchronous Meetings to Asynchronous Collaboration}

\author{Ruotong Wang}
\email{ruotongw@cs.washington.edu}
\affiliation{%
  \institution{University of Washington}
  \city{Seattle, Washington}
  \country{United States}}
\orcid{0000-0003-0964-6943}

\author{Lin Qiu}
\email{lq9@cs.washington.edu}
\affiliation{%
  \institution{University of Washington}
  \city{Seattle, Washington}
  \country{United States}}
\orcid{0009-0003-2354-1192}

\author{Justin Cranshaw}
\email{justin@cranshaw.me}
\affiliation{%
  \institution{Maestro AI}
  \city{Seattle, Washington}
  \country{United States}}
\orcid{0009-0002-5856-1735}

\author{Amy X. Zhang}
\email{axz@cs.uw.edu}
\affiliation{%
  \institution{University of Washington}
  \city{Seattle, Washington}
  \country{United States}}
\orcid{0000-0001-9462-9835}



\def\authnote{1}
\newcommand{\fixme}[1]{\ifnum\authnote=1{\textcolor{red}{[FIXME: #1]}}\fi}
\newcommand{\rw}[1]{{\textcolor{blue}{#1}}}

\begin{abstract}
A recent surge in remote meetings has led to complaints of  “Zoom fatigue” and “collaboration overload,” negatively impacting worker productivity and well-being. One way to alleviate the burden of meetings is to de-emphasize their synchronous participation by shifting work to and enabling sensemaking during post-meeting asynchronous activities. Towards this goal, we propose the design concept of \textit{meeting bridges}, or information artifacts that can encapsulate meeting information towards bridging to and facilitating post-meeting activities. Through 13 interviews and a survey of 198 information workers, we learn how people use online meeting information after meetings are over, finding five main uses: as an archive, as task reminders, to onboard or support inclusion, for group sensemaking, and as a launching point for follow-on collaboration. However, we also find that current common meeting artifacts, such as notes and recordings, present challenges in serving as meeting bridges. After conducting co-design sessions with 16 participants, we distill key principles for the design of meeting bridges to optimally support asynchronous collaboration goals. Overall, our findings point to the opportunity of designing information artifacts that not only support users to access but also continue to transform and engage in meeting information post-meeting. 

\end{abstract}

\maketitle

\section{Introduction}
As work-based collaborations increasingly occur remotely, people are spending more time in online meetings~\cite{covid, augstein2023remote, work}. A 2021 survey reveals that the time information workers spend on voice and video calls has doubled from 7 to 14 hours since the 2020 onset of the pandemic~\cite{hbr}, and the change persisted even with the more recent shift to hybrid work~\cite{bloom2022hybrid}. While synchronous meetings play an important role in supporting team planning, coordination, and decision making, the sharp increase in meeting load has led to problems such as ``Zoom fatigue''~\cite{fosslien2020combat}, where prolonged videoconferencing becomes physically and mentally exhausting, and ``collaboration overload''~\cite{hbr}, where individual task accomplishment suffers due to collaborative group activities. These phenomena have been shown to negatively impact worker productivity (e.g., ~\cite{baym2021collaboration, hbr, nesher2022understanding}) and well-being (e.g.,~\cite{baym2021collaboration, hbr, elbogen2022national, tan2022zoom}).

\begin{figure}
  \includegraphics[width=\linewidth]{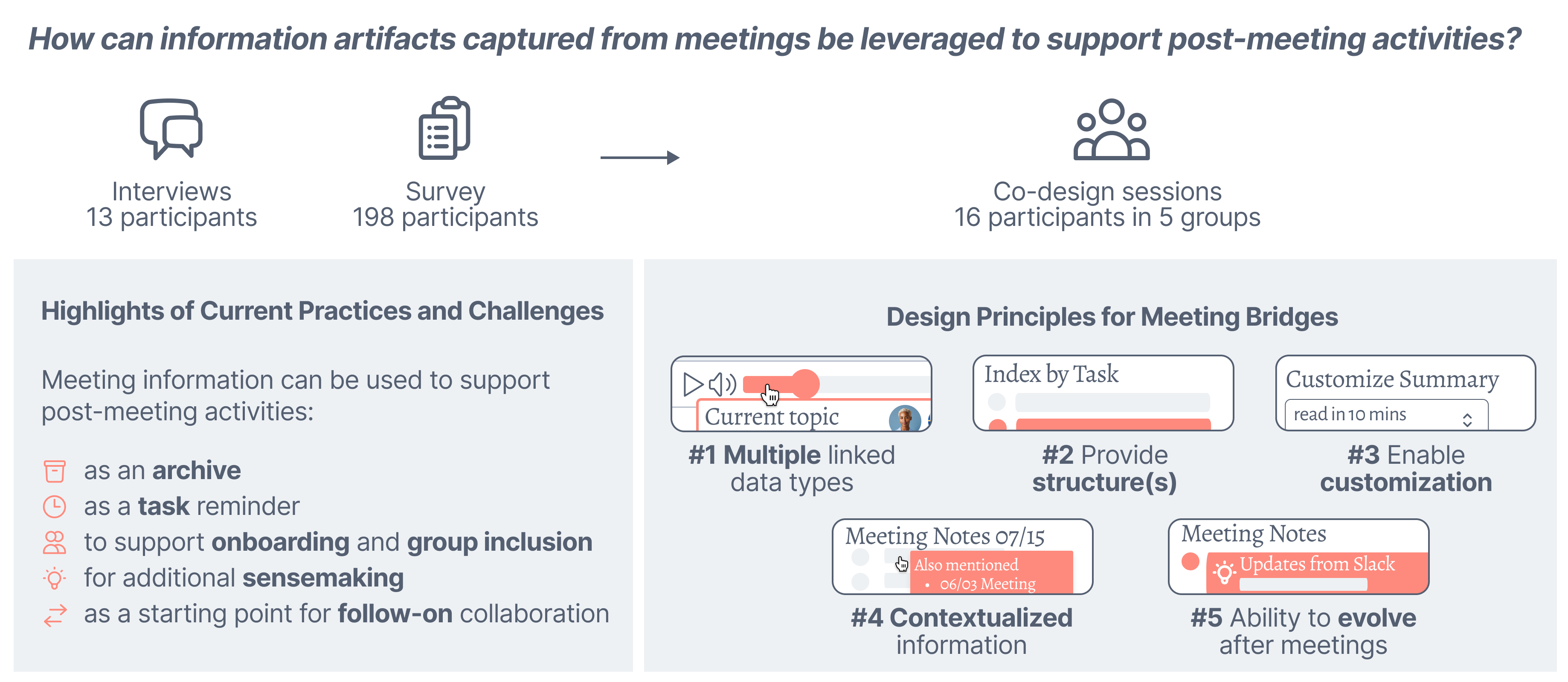}
    \caption{Based on insights from an interview and a survey study on people's current practices and challenges in using meeting information after meetings end, we propose the design concept of \textit{meeting bridges}. Through co-design sessions, we distilled five key design dimensions and created example mockups of meeting bridges to illustrate the opportunity of designing information artifacts that not
only support users to access but also continue to transform and engage in meeting information post-meeting. }
    \label{fig:abstract}
\end{figure}

To achieve a more productive and healthier practice of remote work, many organizations are developing practices to shorten meetings and make them more intentional~\cite{cao2021large} or exploring strategies to eliminate meetings altogether, such as no-meeting days~\cite{cao2021large, laker2022surprising, butler2021challenges}. Superficially, these strategies can reduce time spent in online meetings, but they often do so at the cost of shifting more work to periods before and after the meetings themselves. Studies show that online workplace collaboration significantly shifted the work burden from synchronous communication to asynchronous or decentralized work during the pandemic~\cite{yang2022effects, cai2021understanding, kalmar2022covid}. Indeed, in a survey with Microsoft employees, over half of the participants report spending more time preparing for meetings beforehand---via pre-reads, agenda and goal setting, asynchronous and synchronous discussions~\cite{baym2021collaboration}---than they did pre-pandemic. The same survey shows that follow-ups to synchronous meetings are now prevalent among workers: a vast majority (94\%) spend more time now than previously on post-meeting follow-up activities, via either chats or additional meetings~\cite{baym2021collaboration}. 

Given the prevalence of asynchronous work that occurs \textit{around} meetings, we posit that the opportunity to alleviate the burden of meetings may not lie in simply reducing meeting time, but also in supporting users to more effectively bridge from synchronous remote meetings to asynchronous activities. Inspired by both classical and recent CSCW works that highlight the important role that digital artifacts play in facilitating team coordination and communication~\cite{augstein2023remote, fang2022knowledge, strauss1988articulation, suchman1996supporting, lee2007boundary, bowker2000sorting}, we focus on exploring how~\textit{enhanced information artifacts captured from meetings can be best leveraged to support post-meeting activities.}

We launched our inquiry by identifying U.S. information workers' current practices of using information captured \textit{during} remote meetings \textit{after} the meeting is over. We interviewed 13 people and surveyed 198 additional individuals  
who frequently participate in remote meetings. We find that attendees use meeting information in five key ways after meetings end:  as an archive, as task reminders, to onboard or support inclusion, for additional sensemaking, and as a launching point for follow-on collaboration. However, we also find that existing information artifacts captured from meetings,  such as notes and recordings, have drawbacks when used to fill these post-meeting roles,  
including a lack of support for lightweight but rich ways to capture meeting information, post-meeting difficulties in understanding and filtering relevant information, and hesitancy to share meeting content with others due to trust and privacy concerns. 

Our results highlight an opportunity that informs our basic research question: \textbf{can new artifacts from synchronous meetings be designed to more effectively support asynchronous collaboration?} 
These \textit{meeting bridges} could simplify team efforts to synthesize meeting content and engage with the content in post-meeting activities, regardless of whether a team member participated in the synchronous meeting. To explore the design space of meeting bridges, we conducted a co-design study with 5 groups of 16 U.S. participants. We then distilled five design dimensions and created example mockups of meeting bridges that enable participants to asynchronously explore and make sense of rich content from meetings, share customized meeting summaries with others, and continue to update the artifact as collaboration evolves. We show an overview of our methods and findings in Figure~\ref{fig:abstract}.
We conclude by discussing how existing artifacts map to the design dimensions, how our characterization of meeting bridges could prove useful in generating additional design ideas for tools to support remote collaboration, and the feasibility of implementing these design ideas in real systems. 

Overall, our work makes the following contributions to prior literature: 
(1) we contribute empirical observations that clarify users' needs and challenges in revisiting and engaging in meeting information beyond the duration of remote meetings; (2) we provide a set of design dimensions of meeting bridges, highlighting the opportunity to support users in customizing, updating, and sharing meeting information after the meetings. To show how one can apply our findings, we provide interface mockups that exemplify the possibilities in this new design space, providing a timely answer to researchers' call to extend meetings beyond ``a single point in time''~\cite{WorkTrendIndex}.

\section{Related Work}
\subsection{Recent Shifts in Remote Work}

In response to issues such as Zoom fatigue~\cite{fosslien2020combat} and meeting overload~\cite{hbr}, researchers are calling for more intentional use of meetings~\cite{riedl2022stress, doring2022videoconference, lakerDearManagerYou2022} and a reduction of unnecessary meetings~\cite{cao2021large, laker2022surprising, butler2021challenges} in remote work. 
However, they also warn that merely restricting meetings could hinder employees' opportunities to connect and socialize~\cite{mok2023challenging} and make workers fear being ``left behind''  when not participating in meetings~\cite{WorkTrendIndex}. 

As a result, much work has highlighted the potential of augmenting synchronous meetings with asynchronous communication, such as chat workspaces or asynchronous brainstorming tools, to make meetings more efficient and let workers choose when to engage~\cite{riedl2017teams, stray2021using, shi2021first, macneeBuildingConnectedOrganization2021, lakerDearManagerYou2022}. Indeed, research shows that the shift to remote work has increased asynchronous modes of communication, such as emails and chat messages~\cite{yang2022effects}, and many workers perceive asynchronous communication, such as chat messages, as a positive way to follow up after meetings~\cite{baym2021collaboration}. To support the shifting patterns of communication in remote work, some researchers propose viewing meetings as a \textit{digital artifact} that persists beyond a single point in time, allowing workers to engage when it best suits their needs~\cite{WorkTrendIndex}.

Building on the trend of supporting asynchronous activities that bracket meetings, we conduct a survey and interview study to explore users' current practices and challenges in engaging with meeting information beyond the meeting itself. Our empirical observation that current tools fall short in supporting post-meeting user needs also led us to explore design solutions that support a smooth transition from remote meetings to asynchronous follow-up activities.

\subsection{Supporting Articulation Work in Remote Collaboration}
CSCW has long studied systems and artifacts that support information exchange as part of group collaboration. A central goal has been to establish common ground among group members so that they can coordinate, cooperate, and communicate with each other~\cite{olson2000distance}, which is especially important for tightly coupled collaboration~\cite{marlow2017communication}. 

Team communication usually falls in the category of ``articulation work,'' defined as the necessary work that unites distributed yet interdependent work elements~\cite{strauss1988articulation, suchman1996supporting, star1999layers}. Articulation work  
manifests as tasks such as ``ensuring the flow of resources,''  ``making arrangements about the division of labor,'' or ``linking or meshing otherwise divided tasks''~\cite{strauss1988articulation, star1999layers, pallesen2018articulation}. These tasks, though less visible, are indispensable to collaborative work~\cite{sawyer2006always}. Moreover, the need for articulation work increases as work becomes more complex due to new tasks and technologies~\cite{sawyer2006always}. 

Researchers have explored different ways to design CSCW systems to support articulation work. Some focus on avoiding articulation needs by keeping work separated~\cite{holten2010coordination}; others aim to facilitate the articulation processes by introducing standardization and formal structures~\cite{suchman1996supporting, schmidt1992taking}. For example, Schimidt and Bannon proposed to manage shared information space using a common set of curatable and accessible information objects~\cite{schmidt1992taking}.
Indeed, standardization that is encapsulated in some way (e.g., protocols, procedures, schemes) offers a useful mechanism to reduce the complexity of the articulation process~\cite{bowker2000sorting, ackerman1999organizational}. 

In the context of remote work, meetings can be considered a type of articulation work because they are often used to manage the ``overhead'' of coordinating interdependent work and engaging with  work artifacts from different group members~\cite{esbensen2014routine, raposo2004combining}. Like other articulation work, meetings are indispensable to successful collaborative work. Moreover, with the rise of remote work characterized by greater temporal flexibility, growing complexity, and higher interdependence of projects, articulation work such as meetings and other communication pathways could consume an increasing proportion of worker time, e.g., in assuring team members are on the same page and that primary tasks go well, diverting them from their primary responsibilities~\cite{hbr}. 

Consistent with classical CSCW literature that suggests supporting articulation work using standardization and formal structures, a recent empirical study on two dispersed virtual teams in a global Fortune 500 company showed that digital artifacts such as emails, documents, and webpages play important roles in facilitating the \textit{transfer} (i.e., presenting and accessing artifacts without changes), \textit{translation} (i.e., adding to artifacts while preserving their original state), \textit{transformation} (i.e., molding and changing artifacts), and \textit{holding }(i.e., restricting others’ access to artifacts) of knowledge~\cite{fang2022knowledge}.

These findings highlight the potential of using digital artifacts to facilitate knowledge coordination in virtual teams. However, it is unclear how such artifacts could support the specific information flow for remote meetings. 
The abundance of multi-media information generated by the surge of remote video meetings presents a unique opportunity to design effective digital artifacts for this purpose, thereby supporting the articulation work that is necessary for successful collaboration.

\subsection{Supporting the Use of Meeting Information after Meetings}
Online meetings are indispensable for remote team communication, coordination, and decision-making. While an abundance of CSCW work focuses on supporting the in-meeting experience to make it more productive~\cite{liu2019notestruct, sonItOkayBe2023, mcgregorMoreMeetingsChallenges2017, chandrasegaran2019talktraces}, effective~\cite{murali2021affectivespotlight, miller2021meeting, aseniero2020meetcues} and equitable~\cite{kaminski2016learning, samroseMeetingCoachIntelligentDashboard2021, houtti2023all, CoCoCollaborationCoach},
 recent research on recurring meetings urges a broader approach that considers users' personal need to revisit and continue to engage in asynchronous meeting conversations even after meetings end~\cite{niemantsverdrietRecurringMeetingsExperiential2017}. 

However, rich and unstructured multi-media data captured from remote meetings are challenging for users to consume afterward. 
To address this, many researchers have investigated ways to support users in accessing information captured from meetings. 
Tucker and Whittaker categorized systems into four groups:  \textit{audio} (e.g., an audio browsing tool with speaker segmentation~\cite{kimber1995speaker} and a group of audio interfaces with diverse visual indices~\cite{hindus1992ubiquitous}), \textit{video} (e.g., VideoManga, which generates pictorial video summaries~\cite{uchihashiVideoMangaGenerating1999} and a video browser with two indexes for navigation~\cite{foote1998intelligent}), \textit{artifact} (e.g., Classroom 2000, which combines multimedia artifacts for revisits of lectures~\cite{abowd1999classroom} and Teamspace, an integrated space to access all meeting materials~\cite{geyerTeamCollaborationSpace2001}), and \textit{discourse} (e.g., Rough `n' Ready, which supports named entity search of meeting transcripts~\cite{colbath2000spoken}, and an interactive document generated from captured media~\cite{vega-oliverosThisConversationWill2010}). More recently, researchers have begun to explore systems for richer and more curated media types, such as dashboards (e.g., MeetingVis~\cite{shi2018meetingvis} and EMODASH~\cite{ez-zaouiaEmodashDashboardSupporting2020}, which visualize meeting dynamics to encourage retrospective awareness of emotions). 

These systems commonly support users accessing meeting information by leveraging automatically generated or manually added indexes as structures. Tucker and Whittaker summarized the expected use case of existing systems as \textit{index-centric random access} of meeting information~\cite{tuckerAccessingMultimodalMeeting2005}. The systems assume the need to access information, supporting users in preserving information from meetings and recalling such information after meetings. However, this assumption can be limiting since meetings could extend beyond \textit{a single point in time}~\cite{WorkTrendIndex}. As Fang et al. observe, information artifacts could support not only the transfer of information, but the translation, transformation, and holding of information~\cite{fang2022knowledge}. However, scant work has explored people's needs to mold and transform meeting information to fit different contexts in remote work. One exception is a study on how multimedia artifacts are created, used, and shared in distributed meetings~\cite{marlowTalkingHeadsMultimedia2016}; this work empirically validates users' need to revisit and share media after a meeting and finds challenges, including recording useful information easily and locating scattered information, but solutions to these challenges are still under-explored.

Further, only a few systems explore designs that explicitly support users' continued engagement after meetings other than accessing previously preserved meeting information. Examples include 
HyperMeeting, which allows linking of videos from follow-up meetings~\cite{girgensohn2016guiding}, Video Threads, which supports users to receive video updates for and further contribute to a conversation in the initial meeting~\cite{barksdaleVideoThreadsAsynchronous2012}, and
a post-meeting dashboard that displays meeting information and collects users' feedback for automated meeting understanding~\cite{ehlenMeetingAdjournedOffline2008}. However, these initial explorations tend to focus on novel techniques, instead of generalizable design patterns. 

Building on prior exploration, we investigate users' needs for meeting information in post-meeting follow-up activities. 
To explore concrete designs to address challenges, we conducted co-design studies on information artifacts that could bridge between remote meetings and asynchronous follow-up activities. The results point to concrete design dimensions for digital artifacts that could both preserve information from remote meetings and translate and transform it to fulfill users' subsequent needs.

\section{Methods}
Our interviews and survey of information workers paint a picture of the use of online meeting information in post-meeting activities and the challenges of putting such information to good use. 
Initially, we interviewed 13 participants who frequently participate in remote meetings to assess in-depth their current practices, goals, and challenges when using information after meetings. Preliminary interview themes we uncovered then informed the design of our survey, which included 198 participants. This broader and more diverse sample confirmed our early impressions and helped us to generalize our findings. 
Both studies are exempt from the Institutional Review Board at the University of Washington. 

In both our interview and survey discussions, we scoped our questions about remote meetings to be about small group meetings that occur online via video conferencing tools (e.g., project update meetings, feedback sessions, brainstorming meetings, 1-on-1 check-in meetings). We focus on small group meetings because they are more likely to be a part of a running collaboration that involves follow-on communication as opposed to large lectures or presentations.

\subsection{Interviews} \label{sec:currentMethods}

\textbf{Participants.} We recruited participants online through a screening questionnaire and through word-of-mouth. For online recruitment, we posted the study recruitment message on X (formerly Twitter) and on a 2000-member Slack channel at a large research university in the U.S. We include the screening questionnaire in our supplementary materials.

We began by interviewing local Ph.D. students to quickly gather initial data before branching out to a more diverse pool. Although we strove to capture greater diversity of practices around using meeting information by prioritizing respondents who work in diverse industries, play different roles on teams (e.g., junior, senior), and use varied strategies to capture meeting information based on their answers to our screening questionnaire, our final sample biased towards highly educated individuals in research and the technology industry. We stopped recruiting new participants when in-progress analysis suggested we had reached data saturation.

In total, we conducted 13 interviews (represented by I1--I13).
Participants include 6 Ph.D. students of different stages, 2 designers, 2 research scientists, 1 program manager, 1 professor, and 1 open-source project leader. Detailed participant demographic information and the interview protocol are included in our supplementary materials.

\textbf{Procedure.} Interviews were conducted remotely over Zoom and lasted 52 minutes on average, ranging from 25--66 minutes. All interviews were led by the first author. We started the interviews by asking participants to identify the different scenarios where they needed to keep track of information from remote meetings for potential future usage. For each kind of artifact (e.g., notes and recordings) they mentioned, we asked follow-up questions about their experiences using it. When asking about post-meeting experiences, we focused on specific information participants found useful from meetings, how they wanted to use the artifact post-meeting, and the ways the artifact succeeded or failed to support their collaboration needs.  

\textbf{Analysis.}  
During data collection, the first author debriefed with other authors after each session and kept a running memo to document emerging themes. All authors met weekly to discuss emerging themes. All interview data were transcribed and analyzed via collaborative qualitative coding. The first two authors coded the interview data following guidance from a reflexive thematic analysis approach~\cite{Braun2019} to clarify the practices and challenges of capturing and using such information.

In the first round of coding, the first two authors conducted ``open coding'' on all interview data. Over 500 open codes were generated. All authors then discussed and iteratively clustered the data into common practices and challenges when capturing and using meeting information. Examples of clusters include ``\textit{Create to-do list after meeting from notes and recordings,}'' ``\textit{Use recording to track comprehensive information,}''  and ``\textit{Challenges of locating information from missed meeting.}'' Based on the list of practices and challenges identified, all authors iteratively discussed and derived roles that information from meetings could play post-meeting and challenges in creating, consuming, and sharing meeting information.

\subsection{Survey} \label{sec:currentMethods}
\textbf{Participants.}
We recruited survey participants via Prolific.\footnote{https://www.prolific.co} 
We limited our pool to people based in the U.S., between the ages of 19 and 65, who are employed part-time or full-time, and who have some experience with remote meetings.
Our survey took 12.8 minutes to complete on average. Each participant was paid \$6.04 regardless of their completion time. 
The final survey sample consists of 198 responses; five responses were removed for failing an attention check. Our sample is balanced in gender,  with a 50/50 breakdown, but skewed slightly towards a younger population. It represents participants across a variety of educational backgrounds, industries, company types, and job roles. 
Slightly over half of the participants reported having management experience. We include more specific demographic information in the supplementary materials.

\textbf{Survey Design.}  
Our survey included multiple-choice questions, free-response questions, and 5-point Likert scales. The design was iterated upon and discussed within the research team. We also piloted the survey with 39 students and 3 initial Prolific participants to receive feedback before deploying the survey. We include the survey in our supplementary materials.

We started the survey by asking participants to indicate the information artifacts that they have used to keep track of information in meetings (e.g., using recordings, notes, screenshots). As practices around these different artifacts may vary greatly, and given our survey length constraints, we then asked focused questions on the practices of using \textit{notes} and \textit{recordings}, two widely accessible artifacts that represent distinct media formats and often involve different practices and norms. 

Specifically, for participants using recordings or notes in meetings, we inquired about their frequency of producing and referring to recordings or notes and their likelihood of sharing the produced artifacts with others. We also asked them to mark the goals and strategies they use from a multiple-choice list. We iterated on this list and, and we combined preliminary findings from the interview study and multiple brainstorming sessions with all authors via multiple pilot tests. 

Finally, for all participants, regardless of whether they used recordings or notes in meetings, we asked what could prevent them from using each, as well as their perceptions of the difficulties of completing common tasks using each. We also collected data on participants’ preferences for the type of information they want to capture in remote meetings in different scenarios. At the end of the survey, participants were asked to reflect on their remote meeting experiences in general, regardless of the strategies they selected. 

\textbf{Analysis.}
For multiple choice and Likert scale questions, we analyzed the responses via descriptive analysis. To analyze the open-ended questions, we used inductive coding and clustering to surface common themes in the responses.

\section{Current Practices and Challenges}
\label{info-captured}
\subsection{Current Practices} 
We now describe current practices that interviewees and survey respondents use to capture and revisit information from remote meetings. This includes the type of information they capture, the frequency of using that information in post-meeting activities, and the distinct ways of using the information after meetings.

\subsubsection{\textbf{Type of Information Captured.}} 
\begin{figure}[!htb]
\centering
\begin{subfigure}{.35\textwidth}
  \centering
  \includegraphics[width=\linewidth]{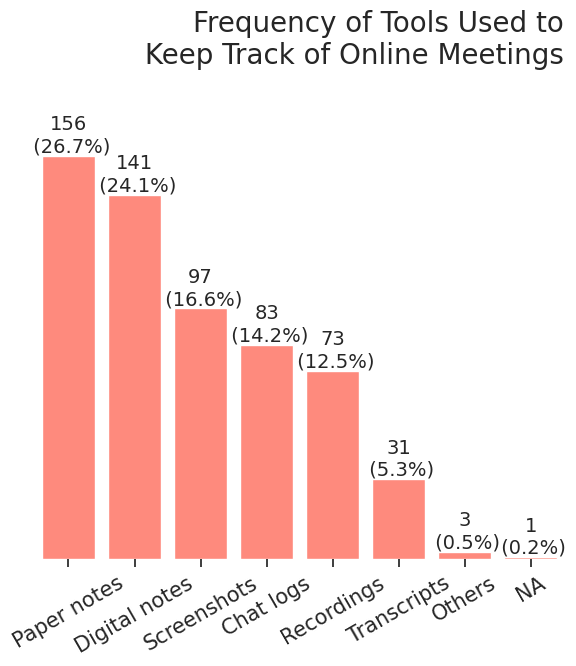}
  \caption{}
  \label{fig:toolFreq}
\end{subfigure}%
\begin{subfigure}{.65\textwidth}
  \centering
  \includegraphics[width=\linewidth]{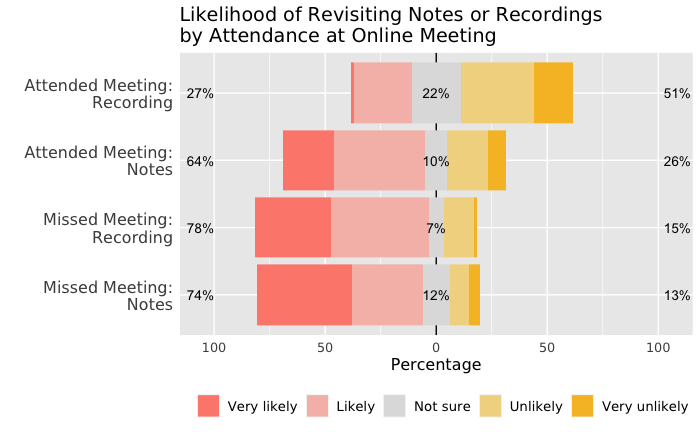}
  \caption{}
  \label{fig:revisitLikelihood}
\end{subfigure}
\caption{(a) The survey result shows that people use a variety of artifacts to keep track of information from meetings. Paper and digital notes are used most widely among participants. (b) Participants reported likely to revisit notes, regardless of whether they attended the meeting or not, but only likely to revisit the recording when they missed the meeting.}
\label{fig:test}
\end{figure}

We found that participants capture information from remote meetings using a variety of artifacts (see Figure~\ref{fig:toolFreq}). 156 survey respondents (79\%) indicated using paper notes, and 141 (71\%) indicated using digital notes to keep track of meeting information. In contrast, recordings were used less frequently, with 36.9\% of survey respondents stating that they keep track of information from remote meetings using video or audio recordings.

For participants who indicated the use of notes and recordings, we additionally asked about the frequency of use for each artifact. Of those who use notes, 63.8\% report using notes in most or almost all meetings they attend. In comparison, only 28.8\% report using recordings in most or almost all meetings. In interviews, participants also mentioned not making frequent use of recordings; those who record meetings usually work regularly in teams that maintain such conventions (I10, I11, I12). For others, the decision to record is case-by-case, either when someone requests it or when the meeting itself is critical and dense with information (I1, I5, I6, I8, I9, I13).

\subsubsection{\textbf{Frequency of Information Use After Meetings. }}We found that while note takers refer back to their notes, those who record meetings rarely do. Most survey participants (80\%) return to at least \textit{some notes} taken either by themselves or others, but only 39\% return to at least \textit{some recordings}. Indeed, most survey respondents refer to only a few recordings (51\%) or do not bother to refer to them at all (9.6\%). This suggests that many recordings taken for synchronous meetings are \textit{never} actually used.  

\begin{figure}[!h]
    \centering
    \includegraphics[width=0.9\textwidth]{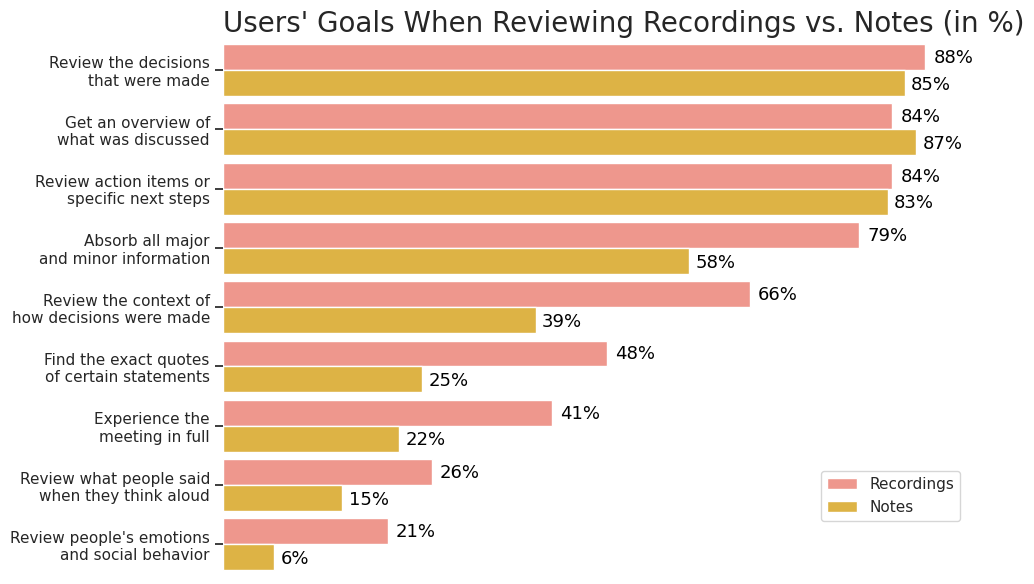}
    \caption{When revisiting recordings, participants reported a more expansive list of goals compared to their goals when revisiting notes.}
    \label{fig:goals}
\end{figure}

Moreover, we asked about the circumstances in which they would refer to recordings or notes after meetings (Figure~\ref{fig:revisitLikelihood}). Over 70\% of survey respondents said they would refer to \textit{notes} if they missed a meeting. When they attended the meeting, 64\% still indicated that they were likely to refer to notes. In contrast, only 27\% indicated their likelihood to revisit \textit{recordings} when they attended the meeting, but 78\% did so when they missed the meeting. Interestingly, when revisiting recordings, survey respondents reported a more expansive list of goals compared to their goals when revisiting notes, including more archival goals, such as: ``Absorb all major and minor information'' and ``Find the exact quotes of certain statements that were said'' (Figure~\ref{fig:goals}).

We also queried how frequently participants would share information captured from meetings with others. 73\% of survey participants reported sharing notes at least sometimes, and 53\% reported sharing recordings at least sometimes after meetings. 

In general, data collection results show that a variety of information is being captured during online work meetings. People frequently use captured information by revisiting or sharing it after meetings end. In particular, information captured in meeting notes continues to support post-meeting asynchronous activities. However, not all captured information is actively used, e.g., recordings are occasionally never seen again after being captured.

\subsubsection{ \textbf{Post-Meeting Uses for Captured Information.}} Our interview results capture a nuanced picture of how information gleaned from meetings is used to support post-meeting activities. We synthesize five main roles that interviewees expressed for such use. 

\textbf{Personal or Group Archive.}
Most commonly, meeting attendees want to keep a record of meeting information to help them personally or collectively revisit it asynchronously when needed. They expect that the archive could provide a truthful representation of meeting events and decisions. For example, I8 described  notes as ``\textit{more true than my real memory}'' as notes are captured in closer proximity to the events.  
Many users find archived information especially useful when the  meeting involves highly dense, complex information that is difficult to remember or digest in real-time. For example, as a non-native English speaker, I13 revisited recordings because they have difficulty following English content in technical discussions but felt  embarrassed to ask clarifying questions. 
For others, an archive of meetings helps fill in lapses of memory. 
When captured for archiving, information is accessed only on an as-needed basis. For instance, I6 recorded meetings in which they received feedback on their work but many times they ``\textit{didn't feel a need to [go back to watch it]}'' because ``\textit{the feedback I got was actually not that much and very manageable}.'' 
Indeed, many participants referred to recordings and notes as a ``\textit{backup}'' (I4,8),  ``\textit{fall-back mechanisms}'' (I13), or ``\textit{safety nets}'' (I13). 
However, this role is important, as I13 states: ``\textit{It [the need to go back to recordings or notes] doesn't happen a lot, but it's very critical.}

Participants also commented on the trade-offs between different artifacts. They often prioritized  revisiting notes due to relative ease of search and refer to recordings only when notes are insufficient; however, they preferred recordings when they want to access nuance (e.g., verbatim quotations) and minor information, revisit visual aids, or understand the context of decisions that were made. 
For example, I10 watched meeting recordings to capture  ``\textit{the insights that maybe the notes didn't catch}.'' 
Often, users had a sense of whether recordings are needed or not before meetings occur based on the meeting's importance, expected information density, and expected 
level of activity during the meeting. For example, I6 said: ``\textit{I only really personally record meetings if I'm getting feedback on something and I feel like I can't type the feedback quick....to make it useful.}'' 
Interviewees also clarified the distinction between personal and group meeting archives. Many mentioned that they would separate personal notes from shared group notes and treat them differently. For example, I2 described how they use specific terminology that makes sense only to them in their personal notes, but use commonly defined vocabulary in shared notes.

\textbf{Personal or Group Task Reminders.} Information captured in meetings can also be used as a reminder to guide post-meeting tasks. 
Here, interviewees generally talked exclusively about notes. 
Several mentioned notes being ``\textit{action-oriented.}'' For example, I3 described how he would document to-do items in group notes to keep track of tasks: ``\textit{Most of the notes are oriented around providing enough information for the people who are going to do those things to go back and remember what they're supposed to do.}'' 
To-do items also get prioritized when people revisit notes for the purpose of reminders. For example, I12 prioritized the following: ``\textit{What are some of the things that I have to do until we meet next time or whatever the deadline is? What are some of the things that I need from other people? What are some of the things that I need to do for other people? }'' 
Unlike meeting archives, interviewees preferred for reminders to be documented in ways that could be made visible to people assigned to tasks. For example, I6 took  notes in Google Docs because they could tag people with specific tasks.  I6 also liked when people send a post-meeting follow-up email containing a to-do list.  

\textbf{Supporting Group Onboarding and Inclusion.} Often in group collaborations, a member could miss a meeting, or a new member could  join after the start of the project, or the outcome of the meeting must be shared with a broader audience. 
Information captured during meetings helped bring these members onboard. 
Interviewees stated that richer formats  especially help when someone missed a meeting.
For example, I9's team, which does not regularly record team meetings, recorded them when ``\textit{someone really wanted to be at that meeting because they wanted to hear about it but are not able to be there.}'' 
However, those seeking to catch up generally wanted to focus on the main takeaways or decisions derived from the meetings, not on comprehensive details. Especially for meetings deemed relatively unimportant, such as project meetings where they play a supporting role, interviewees found  that ``\textit{there's not a need for keeping track of everything.} (I7)'' 
Some missing content is not necessarily a problem since it  
could serve as an invitation for follow-on communication, which we describe in Section~\ref{sec:followon}.
For example, I11 mentioned that they ask clarifying questions when notes or recording cannot answer all their questions.

\textbf{Additional Sensemaking.} Information captured in meetings can serve as raw materials for meeting attendees to ingest when further sensemaking. One way is by allowing people to make new connections between the information captured and additional information from outside the meeting. For example, I6 usually kept track of their to-do items on a Trello board during project meetings. However, since they were involved in multiple projects, they needed to assess the priority of each item in the context of all projects: ``\textit{[the Trello board] gives that bird's eye view of seeing everything at once... so it really helps me make that kind of assessment when I have them all out in front of me.}'' I11 also explained that looking across several weekly meetings together lets them not only get a general sense of ``\textit{what has been most recently happening,}'' but query more interpretive information, such as ``\textit{the timeline of how did a feature get made}'' in their open source project. 

Additional sensemaking also occurred when participants refer to and re-organize information first captured during a meeting. I7, who is involved in a complex research project, often reorganized their notes of project meetings as they revisited them, and this process helped them further digest the discussion: ``\textit{I'll have written like really shorthand something [during the meeting] like more citations about X. And then maybe I'll go back [to the notes] and specific links or flesh out why we needed to do that [after the meeting].}'' I10, a user researcher, shared that when returning to recordings of meetings with customers, they looked for
``\textit{their interpretation of what they've heard, or [a] colleague's interpretation}'' to make sense of what occurred in concert with others.

\textbf{Starting Point for Follow-on Collaboration.}~\label{sec:followon} Information captured during meetings also motivates or structures follow-on collaborations. For teams with recurring meetings, captured information provides a structure for subsequent meetings. Several participants (I2,I3,I5,I6,I7) mentioned using dedicated Google Docs to take notes for recurring project meetings. 
For example, I6 shared that: ``\textit{I can look back and be like, oh this is what we talked about last week or these were the action items out of last week. And so it gives you that starting point for the next meeting.}'' 

Follow-on collaboration can include follow-on asynchronous discussions since ``\textit{a ton of stuff could happen between two weekly meetings}'' (I11). For people who attended the meeting,  asynchronous conversations centered  around updates from topics discussed during the meeting. For example, I7 posted updates to to-do items from weekly meetings during the week and solicited additional feedback: ``\textit{I'll provide an update on Slack [for the to-do items]. For example, when I make a graph, [the other collaborators] will say oh this is pretty good, but actually, we should reorganize or change these colors or make other smaller adjustments. And then I'll update again in Slack later.}'' I12 noted that they would add questions that they were unable to clarify on the spot to a shared feedback document for people to discuss asynchronously. 

For those not present at the meeting, information captured during the meeting offered an entry point for them to contribute to the conversation asynchronously. For example, when I2 went through notes from meetings they missed, they  saw the process as a ``\textit{simulation of the meeting by imagining that they were there}'' and considered any disagreements with the decisions made. If there was a decision they did not understand where  ``\textit{[their] input would have changed the decision or added more background or context to a piece of decision,}'' they would initiate a conversation with meeting attendees to revisit the decision. Similarly, I11 found it challenging to participate in every meeting synchronously, but they found that reviewing meeting notes lets them participate in meetings asynchronously: ``\textit{Someone had a meeting last week that I wasn't in, but they wanted my opinion on something. So they linked me to the meeting notes and Slacked me...The question was very easy...so I just gave them my answer.}'' 
However, when questions become too complicated to discuss asynchronously, interviewees would defer it to the next synchronous meeting. Indeed, I12 observed how looking back to meeting notes from a missed meeting helped them more easily continue the conversation with the team in a follow-on meeting.

\subsection{Challenges } 
Despite the benefits of meeting information in supporting post-meeting collaboration, we observed challenges for those trying to create and use such information with current tooling. 
We separate challenges according to three stages of the meeting information lifecycle: \textit{capturing} information during a meeting, \textit{consuming} information after a meeting, and \textit{sharing} information with others.

\subsubsection{\textbf{Challenges with Capturing Information during Meetings}}\hfill

\textbf{Lack of tools to capture rich data in real time.}
Participants in our survey and interviews reported being unable to write or type quickly enough to follow meeting progress, resulting in capturing only partial notes during meetings. In the survey, 74\% reported that they \textit{often} or \textit{always} have to \textit{take notes in short phrases/incomplete sentences during meetings}. I3 explained the challenge this way: ``\textit{I just capture random words that people say... 
not always coherent to other people [who read the notes].}'' I13 echoed that their notes often include only partial information, such as the outcome of a discussion, not the nuances, such as ``\textit{justification or trajectory that led to decisions}.''

The tools used to capture information also constrain what they can keep track of. For example, I12 found it challenging to keep track of digital links when taking notes on paper. However, I13 took notes on paper or tablet instead of digital documents so they could draw diagrams to represent ideas non-linearly. 
To address this limitation, some participants used multiple tools for information capture, such as recording content as backup so they can ``\textit{relax a little bit in my note-taking}'' (I10). Participants referred back to the recording to complete a more detailed version of their notes or confirm the notes' information. I11 said: ``\textit{The meeting minutes are great, [but] it's likely missing context. There are nuances to the discussion that aren't contained...}''

\textbf{Capturing information affects meeting engagement.}
Of the respondents who took notes, 82.9\% reported doing so themselves in some or many of their meetings. 
I8 took notes by trying ``\textit{...to type as the other person talking [sic] or as I'm talking.}''  However, note-taking can be ``\textit{cognitively demanding,}'' as indicated by I5 and many others, because it usually involves a lot of ``\textit{context-switching...not only the tools' context but also my mind's context} ''(I5). As a result, note-taking often distracted attendees from actively participating in meetings.

While recordings captures rich data and also makes few cognitive demands compared to notetaking, it still presents challenges for meeting participants. 
83 survey respondents indicated that ``Participants might be hesitant about speaking up if it's recorded.'' A similar concern was echoed by our interviewees. For instance, I12 observed that ``\textit{People become more conservative about the comments that they make because they know it's being recorded...}'' I13 explained that  ``\textit{People could be suspicious about why you are recording. Like are you gonna use this recording against me in the future, you know, when something bad happens?}'' Furthermore, participants mentioned that getting explicit consent from all attendees could be awkward given that individual preference for being recorded or documented might not be explicit during meetings.

\subsubsection{\textbf{Challenges with Consuming Information after Meetings}}\hfill

\textbf{Lacking the necessary context for understanding.}
Despite being  prevalent, notes lack critical information to be understandable to all participants for post-meeting use. For example,  I3 found it challenging to catch up on missed meetings by  reviewing notes taken by other people because they lack sufficient context to follow others' thought processes. Additionally, note-takers often used  abbreviations or notations that others might not understand, or even they themselves may forget after a period of time. 
Even meeting transcripts, which capture conversations more comprehensively, can be insufficient in providing context for meetings that involve sharing of multimedia information. For instance, though revisiting meeting transcripts after meetings worked well for I6, it proved ineffective for design critique meetings, which involve extensive discussions of visual elements: ``\textit{People use words like `it's this' or `move that there,' but I don't necessarily remember the context here and I have no idea what they're talking about.}'' I13 tried to revisit transcripts after meetings but found them unhelpful because the automated transcript incorrectly transcribed foreign names or technical terms.
Several participants referred to information captured in multiple media formats when they found an individual media format to be insufficient. 
For instance, I2 spent extra time returning to more raw sources, such as the recordings, to recover context not captured in notes.

\textbf{Difficulty locating desired information.}
Raw information captured from meetings can be noisy, and people need to filter it to find useful content. However, many found that this process is burdensome. Per I2: ``\textit{The meeting takes an hour, but it takes almost an hour to go back through [the] recording.''} 
I8 agreed, especially for transcripts that ``\textit{...are too long and hard to read.}'' I9's team used to record every meeting for archival purposes but soon abandoned the practice since she found that people rarely return to the recordings, and it became ``\textit{more than anybody can handle}.'' 
I3 found it hard to quickly catch up with missed meetings using recordings because lab meetings often involve ``\textit{random shoot the breeze kind of talking}.''
I1 preferred notes to quickly get a sense of what happened during meetings because they ``\textit{filter out irrelevant information.}''

When seeking specific information, participants reported challenges in locating it. For example, I11 found it difficult to search for information in video or audio recordings: ``\textit{I wish I could jump to a point in the city council meeting where a bill was being discussed. The city council meetings are two hours long, and I don't want to sit there and scramble through the timeline and find that point.}'' Even when functions such as search via keywords are available, I13 still found locating specific information to be challenging: ``\textit{Maybe I don't know the terms to search for [desired information]... Maybe I remember a scene in a video but I don't know how to search for the scene. Like I need to know the meta information to be able to find it.}''

\subsubsection{\textbf{Challenges with Sharing Information with Others}}\hfill

\textbf{Reluctance to share personal notes or to trust notes taken by others.}
Some participants avoid sharing notes due to the personal decisions they made regarding both format and content while note-taking.
One survey respondent said: ``\textit{I generally do not share my informal notes from meetings, as those are structured for my own purposes.}'' Indeed, we discovered that participants had highly personalized note-taking behaviors, including different forms of organization and structure. They also often prioritized what to capture based on their personal needs for revisiting meeting content. One respondent thought that others would not be interested in their notes: ``\textit{Nothing seemed important for everyone to know. Sometimes it's just things I want to keep to myself to help with my work.}'' Another expressed a lack of confidence in the quality of their notes.

Even when shared, some participants expressed hesitation about trusting notes taken by others, instead preferring to take their own. For example, I2 worried about the mismatch in their judgment of relevancy: ``\textit{they might write things that I don't care about or they might not write things that I care about.}'' Similarly, I10 said: ``\textit{When I have a colleague help me take notes...they're not as invested in that conversation as I am because this is going to be informing my work. It's just a trust issue and I don't trust that they're going to catch the things that I need to catch}.'' Others expressed concerns over being misled by biased interpretations of team member notes. For instance, I8 often used notes as an archive for meeting information, but ``\textit{A lot of people put their thoughts in their notes, which for me, compromises [notes] as a source of memory.}'' I10 echoed the concern: ``\textit{People who create action items [in notes] might have their own bias on what's important and what's not important... I won't be able to form my own opinions.}'' 

\textbf{Reluctance to share recordings due to privacy concerns.}
Privacy of documented information is another common concern that prevents the sharing of meeting content, particularly for recorded data. Only 53.4\% of survey respondents indicated having shared their recordings with others at least occasionally. When asked open-ended questions about why, respondents expressed concerns regarding privacy, namely, that it would be difficult to edit out sensitive information before sharing. This was also reflected in survey question results probing the difficulty of performing certain tasks when using recordings; the task rated most difficult was ``Share only selected portions of a meeting with other people,'' with 38\% of respondents finding it difficult or very difficult.

\section{Meeting Bridges: Designing information artifacts for use after meetings}\label{design-pattern}

Our interviews and surveys highlight that a variety of information is being captured in the form of recordings and notes during remote meetings. This information has the potential to facilitate post-meeting activities. However, much of it is uncurated or poorly curated, making it difficult to use. This presents a significant challenge in forming a critical bridge from remote meetings to post-meeting collaboration.
To address this challenge, we take inspiration from prior work on boundary objects and articulation work in CSCW~\cite{augstein2023remote, fang2022knowledge, strauss1988articulation, suchman1996supporting, lee2007boundary, bowker2000sorting} that inform the design and use of information artifacts for providing a shared structure and space for team members to negotiate and coordinate with each other. 
We explore the potential of designing enhanced \textit{information artifacts} captured from remote meetings that could more effectively serve the need for asynchronous follow-up work after meetings.  

We refer to curated information artifacts that are explicitly designed to support bridging from  meetings to post-meeting  collaboration as \textit{meeting bridges}. \textbf{A meeting bridge is a type of information artifact created in a synchronous meeting with the explicit purpose of facilitating or bridging to asynchronous interaction and collaboration after the meeting is over.} For example, a video recording of an online meeting is a common type of meeting bridge since it is created during a synchronous meeting with the intent of being useful for interaction and collaboration post meeting.
Our distinction of which artifacts constitute meeting bridges rests on the \textit{intent} being about supporting post-meeting activity. Thus, throwaway information for facilitating ongoing synchronous interaction, such as a chat backchannel or sketches on a whiteboard, is not intended for bridging; however, this content could get \textit{repurposed} into an artifact that does seek to serve post-meeting bridging aims; e.g., live captions of the dialogue during a meeting could also be incorporated into a meeting transcript for people to read and review afterward. 

We call attention to the design concept of \textit{meeting bridges} to highlight opportunities in designing information artifacts to better support bridging. However, the concept also gives rise to a set of new questions. In particular, how can designers concretely approach the design concept? What kind of formal structure should a  meeting bridge embody to reduce redundant articulation work? 
To explore the design space of meeting bridges, we conducted a co-design study with five groups of 16 total participants to elicit design ideas. We first describe our study methods (Section~\ref{designMethod}) and then discuss the design dimensions we synthesized from the study (Sections~\ref{sec:DD1} to ~\ref{sec:DD5}).

\subsection{Co-design Study Methods}~\label{designMethod}
We conducted co-design sessions to elicit design ideas from participants for \textit{meeting bridges}. These sessions also sought to elicit our participants' 
desires for different forms of meeting bridges.
We conducted 16 individual study sessions where we asked participants to sketch desired designs based on real or realistic synchronous meetings that they participated in. To help them explore design ideas more concretely, we collected and made available to them raw information (e.g., recordings, notes, chat logs) captured from synchronous meetings during sketching activities. The co-design study is also exempt from the Institutional Review Board at the University of Washington.

\textbf{Participants.} Given that our co-design activity uses content produced from real or realistic meetings our participants attended, we recruited participants through word of mouth to reduce  privacy concerns. Participants were recruited in groups of people who knew each other so that they could share a more natural experience in a remote synchronous meeting. We recruited a mix of designers and lay users working in different occupations. In total, 16 participants were recruited and divided into five groups to participate in real or realistic meetings. We include a table of more specific demographic information in our supplementary materials.

\textbf{Preparatory online synchronous meetings.}
Our study design necessitated access to raw materials from a remote meeting.
For groups with a scheduled existing meeting, we asked them to record one session of the meeting and share the recording and any notes with us. 
For groups without an existing meeting, we set up a mock meeting and captured recordings and notes. Of the five groups, 2 recorded existing meetings, and 3 recorded mock meetings. All meetings were conducted over Zoom except for one real meeting, which took place on Google Meet.

For groups that had a mock meeting, we tasked them with coming up with proposals for a fund that is dedicated to improving an aspect of their work environment, adapted to each team's specific context. We designed the task so participants could have open-ended discussions on a general topic relevant to their work. 
Additionally, we selected topics that were information-dense, so participants could hypothetically imagine using the artifacts after the meeting. 
All task prompts are provided in the supplemental material. 

We collected as many forms of rich information artifacts as we could from the mock meetings. We video-recorded the meetings and also asked participants to take notes so they could continue the discussion in a hypothetical follow-up meeting and so a hypothetical team member who missed the meeting could catch up afterward. Two paper coauthors also attended the meetings; during the ensuing discussion, the coauthors were muted and had their video turned off while making detailed logs of notable events (e.g., someone sharing their screen). 
For groups that recorded real meetings, we asked one participant to take group notes and obtained consent from every meeting participant before the meeting to capture a recording. We then produced a log of events by watching the recording obtained from the group.

\textbf{Co-design sessions.}
We conducted 16 individual co-design sessions with each participant  1--3 weeks after the meeting. The sessions were conducted in person to make it easier for participants to sketch and share their ideas. The median length of a session was 90 minutes, with a range of 45--95 minutes. 
There were two components of the co-design session: (1) an \textit{interview} about their previous experiences with meeting information artifacts, followed by (2) \textit{sketching} ideas for meeting bridges from the preparatory meeting. The interview covered questions similar to those posed in Study 1 interviews and was designed to stimulate participant thinking about challenges in their experiences and potential solutions. We include our interview questions in the supplementary materials. 

The main co-design activity was a series of sketching exercises that aimed to elicit participants' ideas about desired characteristics of meeting bridges.  
We introduced 5 scenarios that covered different ways in which information artifacts from meetings could be used. We also adapted the scenarios for each group to be relevant to their specific discussion context. Specifically, participants were asked to sketch designs of artifacts that could be (1) used for personal revisiting, (2) shared with team members who also attended the meeting, (3) shared with lightly involved team members who missed the meeting, (4) shared with heavily involved team members who missed the meeting, and (5) shared broadly.

\begin{figure}[!tbp]
  \centering
  \subfloat[]{\includegraphics[width=0.45\textwidth]{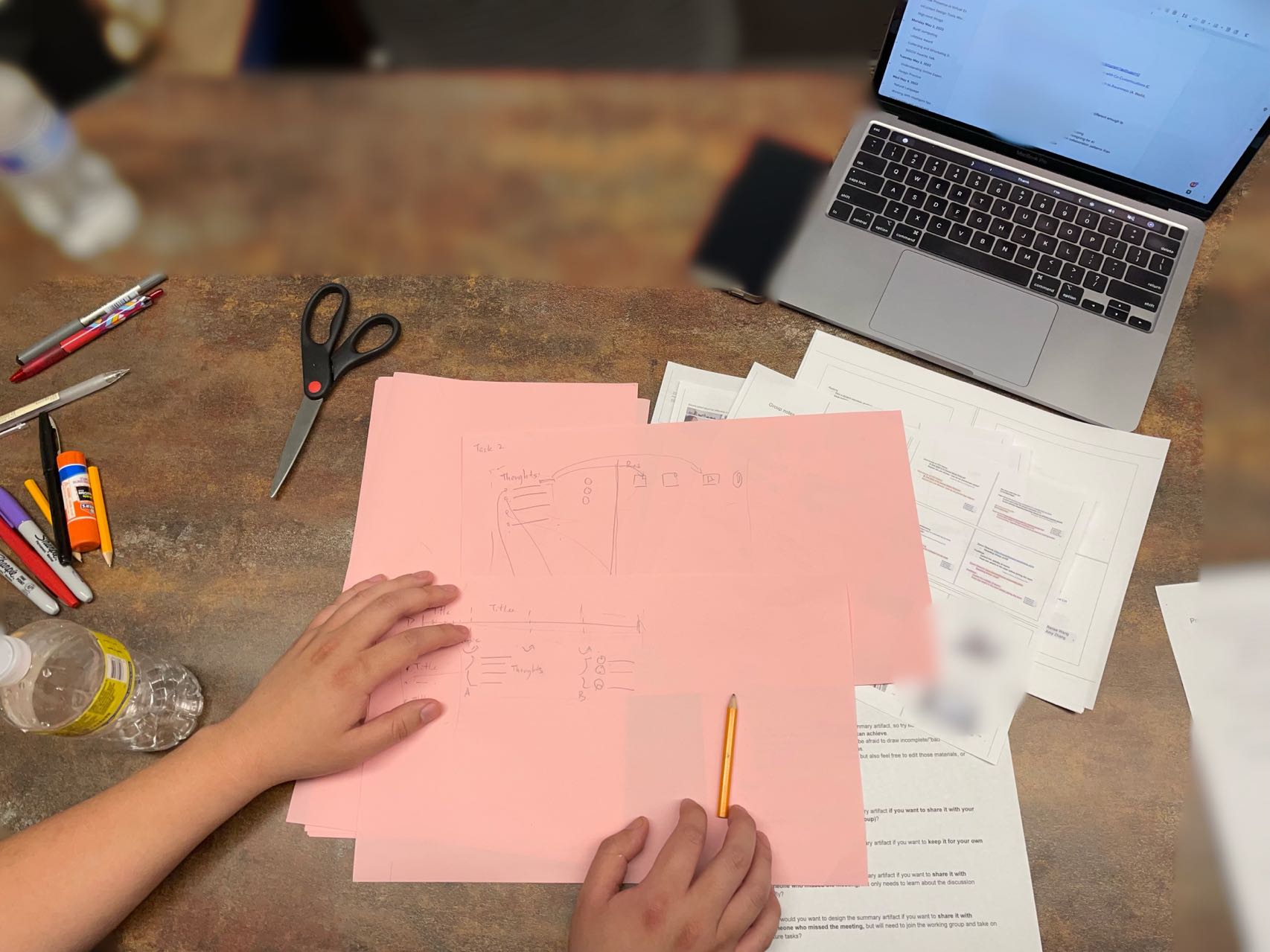}\label{fig:codesign1}}
  \hfill
  \subfloat[]{\includegraphics[width=0.45\textwidth]{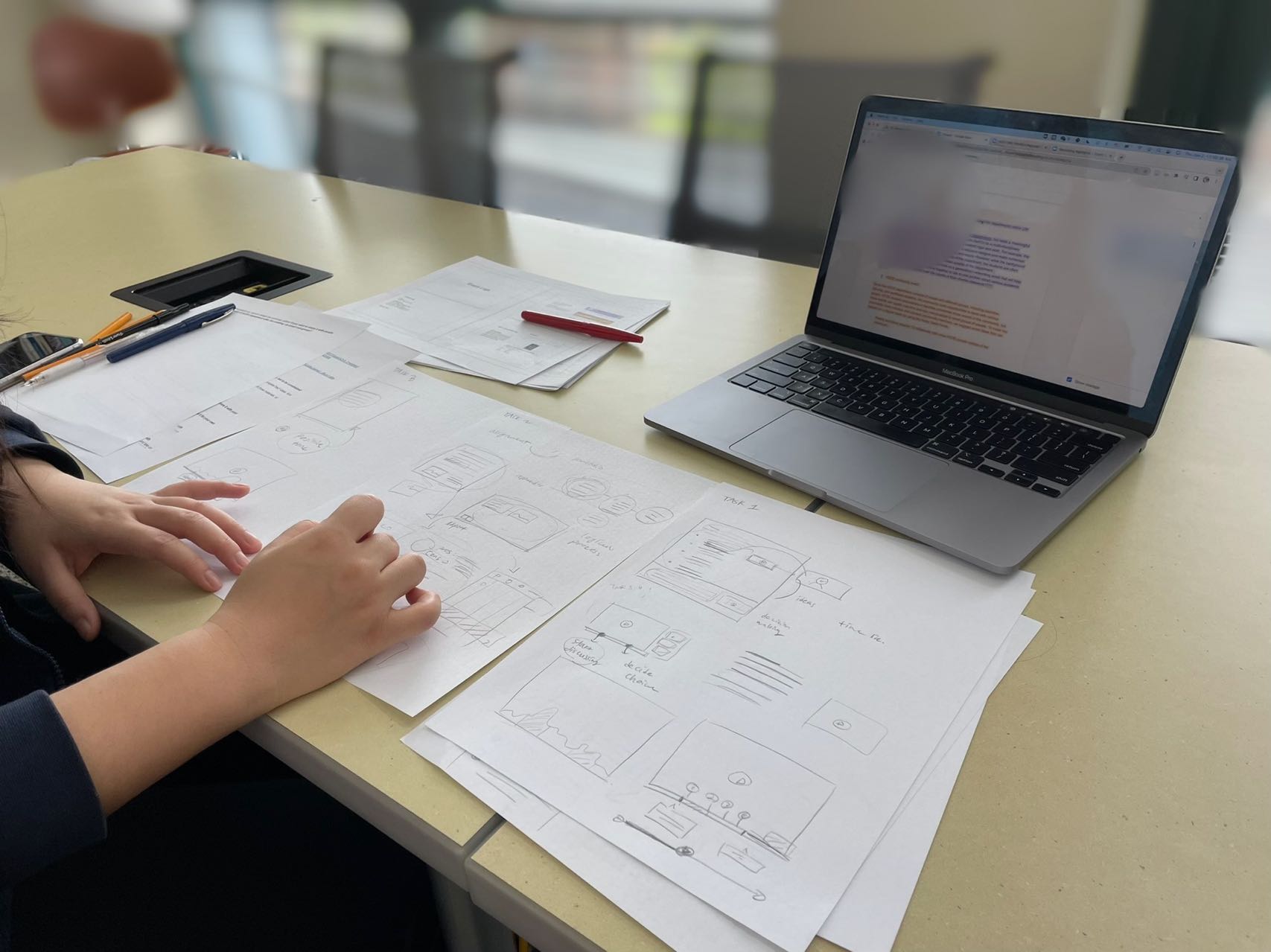}\label{fig:codesign2}}
  \caption{Co-design session setup. Parts of the images are blurred for anonymity. Participants are asked to sketch ideas for meeting bridges from the real or realistic meeting that they had. We provided a variety of existing information artifacts from the meeting (e.g., notes, recordings, transcripts, etc.) as materials for participants.   }
  \label{fig:setup}
\end{figure}

To prepare for the sketching activity, we organized the participant's personal notes, any meeting scribe's notes, our log of events, transcripts of the dialogue, the chat log, and the video recording into one Google Drive folder. We also printed out all artifacts that are text documents as well as screenshots from the video recording. The prepared materials served as prompts for users to consider how their desired artifacts were similar or different from existing ones. These elements could also form ``building blocks'' for participants to consider how to recombine them into a new design. If participants wanted to use any of the printouts, we provided scissors, glue, markers, and highlighters. See Figure~\ref{fig:setup} for pictures of our setup.

We asked participants to sketch on paper and explain their thinking out loud throughout the session. In a few sessions, the coauthors sketched on behalf of participants when they expressed being more comfortable describing their idea in words; in these cases, we confirmed that the sketches aligned with what they imagined. When participants finished sketching for each scenario, we also asked them to describe how they imagine the artifacts being used. This helped us understand users' need for meeting bridges in context. Participants could work on any of the five scenarios in no particular order. In some sessions, participants decided that they wanted similar designs for multiple scenarios. We include example sketches in our supplementary materials.

\textbf{Analysis.}
All co-design sessions were audio and video recorded and then transcribed into textual format. We coded all data using a collaborative coding app. The first two coauthors coded the transcripts along with the sketches following guidance from a reflexive thematic analysis approach \cite{Braun2019} to identify design ideas for artifacts as well as users' needs. Our initial round of coding and clustering generated 12 groups of design patterns. We further discussed these design patterns among all authors and consolidated them into broader themes about desired meeting bridge characteristics.

\subsection{Five Design Principles for Meeting Bridges}
We organize our results around five high level design principles that arose from our analysis. Note that we do not claim these design principles to be comprehensive given that our co-design participants skew towards knowledge-intensive work. However, they provide examples of how we might design artifacts with features that explicitly support bridging.

\begin{figure}
\minipage{1\textwidth}
  \includegraphics[width=\linewidth]{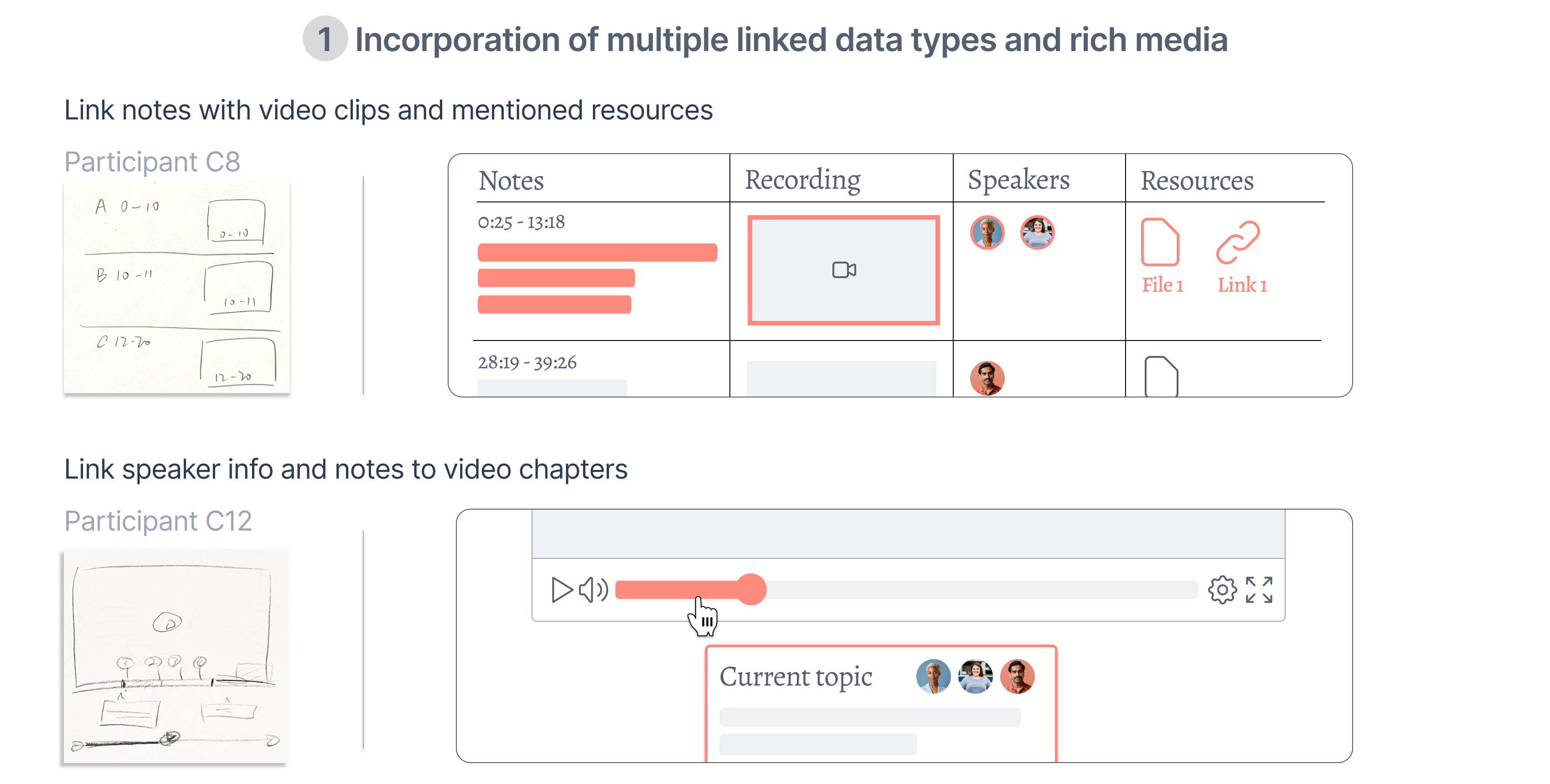}
    \caption{Design Principle \#1: Incorporation of multiple data types and rich media presentations. We show two design examples: link notes with video clips and resources (top) and link speaker information and note snippets to video chapters (bottom). For each example, we show original sketches collected in co-design sessions, as well as a synthesized mockup based on the sketches.}
    \label{fig:DD1}
\endminipage\hfill
\end{figure}

\paragraph{\textbf{\#1: Incorporation of multiple linked data types and rich media.}}\label{sec:DD1}
Most participants wanted meeting bridges to incorporate multiple data types and media formats (Figure~\ref{fig:DD1}), as this could help users revisit or build on meetings from multiple perspectives. For example, C7 wanted meeting bridges to ``\textit{combine different forms of media to be as comprehensive as possible.}'' When asked about the kind of information to include in the artifact, C14 said ``\textit{every piece of information here,}'' referring to all materials we provided. Participants additionally wanted links shared, attribution of information source (e.g., speakers) and meta-information (e.g., information about meeting participants). 
Meeting bridges with multiple data types offered a more comprehensive, nuanced, and accurate view of the meetings. 
For example, per C9, text notes sufficed for understanding meeting outcomes at a high level, but recordings were necessary to grasp the discussions leading to the outcomes. 
Incorporating various data types also enabled individuals to access and make sense of the information using multiple senses, enhancing accessibility per C4, C7, and C11.
However, there were some use cases when comprehensive information was less desirable. For example, C13 noted that meeting bridges, when serving as task reminders, should provide just enough information for people to recognize the content without going into comprehensive detail:``\textit{The purpose [of the meeting bridge] is more of a reminder, so if people just see one sentence and they can remember what happened in the meeting, that's enough. If they do forget, they can just go back to see the full log.}''

Moreover, participants also wanted to connect different media formats so they could understand each piece of information in the context of others. 
For instance, almost all participants suggested linking notes to specific time stamps on video recordings. C12 wanted to see version histories of notes aligned with a video timeline for context of how the text notes were created. For C15, recordings also helped him re-experience the meeting:  ``\textit{Sometimes people write down decisions they made, but not the rationale [in notes]. If I don't [initially] like this decision, but I was probably convinced for some reason [during the meeting], but I don't remember why. I kind of want to go back. I want to know how I was convinced.}'' 

Another commonly desired linkage was the attribution of ideas and opinions to specific participants. 
C15 thought that documenting the originator and supporter of ideas discussed in meetings lets him ``\textit{put the rationale into context.}'' This information is especially desirable for brainstorming meetings, where people could hold vastly different opinions. C11 and C16 similarly thought  that knowing who said what in meetings is helpful to understand the ``\textit{weight}'' of opinions, especially when making decisions. 
C3 added that linking people to ideas helps teams follow up with the corresponding person post meeting.
However, participants also raised concerns that attributing ideas to specific people could negatively affect discussion dynamics. For example, C15 explained that it is important to leave ``\textit{gray ideas}'' in the design. Assigning people to ideas that are then placed on opposing sides could force people into separate `camps' when they might have been open to either side. 
Similarly, C16 thought that the linking need not be one-to-one but instead could be a many-to-many connection that showed a high level aggregate sense of who is interested in what.

\begin{figure}
\minipage{1\textwidth}
  \includegraphics[width=\linewidth]{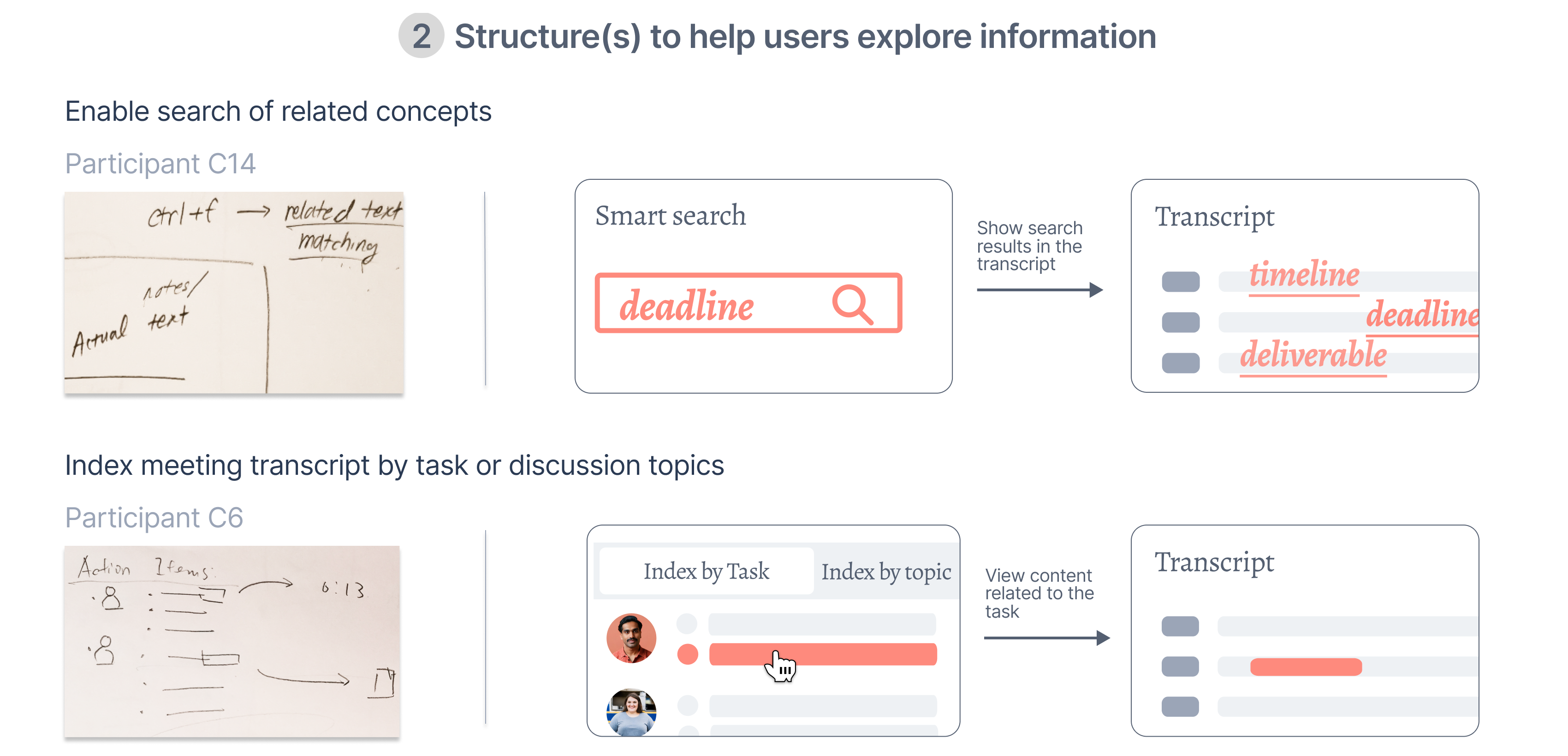}
    \caption{Design Principle \#2: Structure(s) to help users explore information. We show two design examples: enable search of related concepts (top) and index meeting transcripts by task (bottom).}
    \label{fig:DD2}
\endminipage\hfill
\end{figure}

\paragraph{\textbf{\#2: Structure(s) to help users explore information.}} 
Though participants wanted richness, presenting meeting information in its raw form is overwhelming. Several (C4, C11) mentioned that they did not want unedited video recordings or raw notes from synchronous meetings. Instead, they wanted meeting bridges to be ``\textit{skimmable}'' (C14) so they could ``\textit{glean from it}'' (C4). 

Participants proposed several ways of making artifacts more easily consumable. 
Some mentioned adding a summary to help the audience make sense of the rich information. C13 proposed that ``\textit{someone should write a summary [of the meeting] on the document in the form of bullet points.}'' C7, C9, and C14 also proposed one-line or paragraph summaries for main ideas in the meeting. These summaries could serve as a ``\textit{table of contents}'' (C7) for users to explore the content using ``\textit{their preferred way of navigating [information]}'' (C13). 
Participants also wanted to curate and organize information using more explicit  structures. While temporal order is often the default structure for  displaying information, and C6 stated that a temporal index helps him skip to relevant parts, C2 found it not useful as conversations can progress non-linearly. Relevant ideas and comments might not be in temporal proximity, which often led C2 to ``\textit{go through the whole recording [to find what I want].}'' 
Even for notes, C15 found it hard to consume and locate information in a temporal order because ``\textit{it's not a reflection of how [he] thinks about the project}.'' 

Several participants proposed indexing meeting information as a way to create structure that aligns with their goals and mental models. Participants identified a variety of indices, including person, topic, tasks, resources, and action items. 
For example, C16, working on an academic research project, proposed using the academic papers mentioned in the discussion as an index. C6 echoed the idea: ``\textit{I really like to have the thoughts linked to shared paper, video or links. I think this kind of structure would be helpful...quotes are no longer organized by [temporal] order. People are no longer an important part of it. }'' C1 and C3 further proposed to visualize information in the form of a mind map, making it easier to skim.

Finally, some participants proposed searchable artifacts that go beyond the structure enabled by an index. C4 wanted searchable components (e.g., transcripts) so she could search for relevant discussions using keywords:  ``\textit{If I attended the meeting, I always have these little sound nuggets in my head---I know we were talking about something and this person said the word `back and forth'... I attach a lot of my own social-emotional meaning to certain types of words. So in a full transcript...even if it's one little word, I would be able to find that part of the meeting.}''
Some participants desired even more flexibility in search (C6, C14). For example, C14 proposed to not only search for the exact text matching, but to also search semantically related concepts or discussions (Figure~\ref{fig:DD2}): ``\textit{if I were to type `writing' , `reading,' something that's semantically related would populate there. That's really important for my skimmability.}''

\begin{figure}
\minipage{1\textwidth}
  \includegraphics[width=\linewidth]{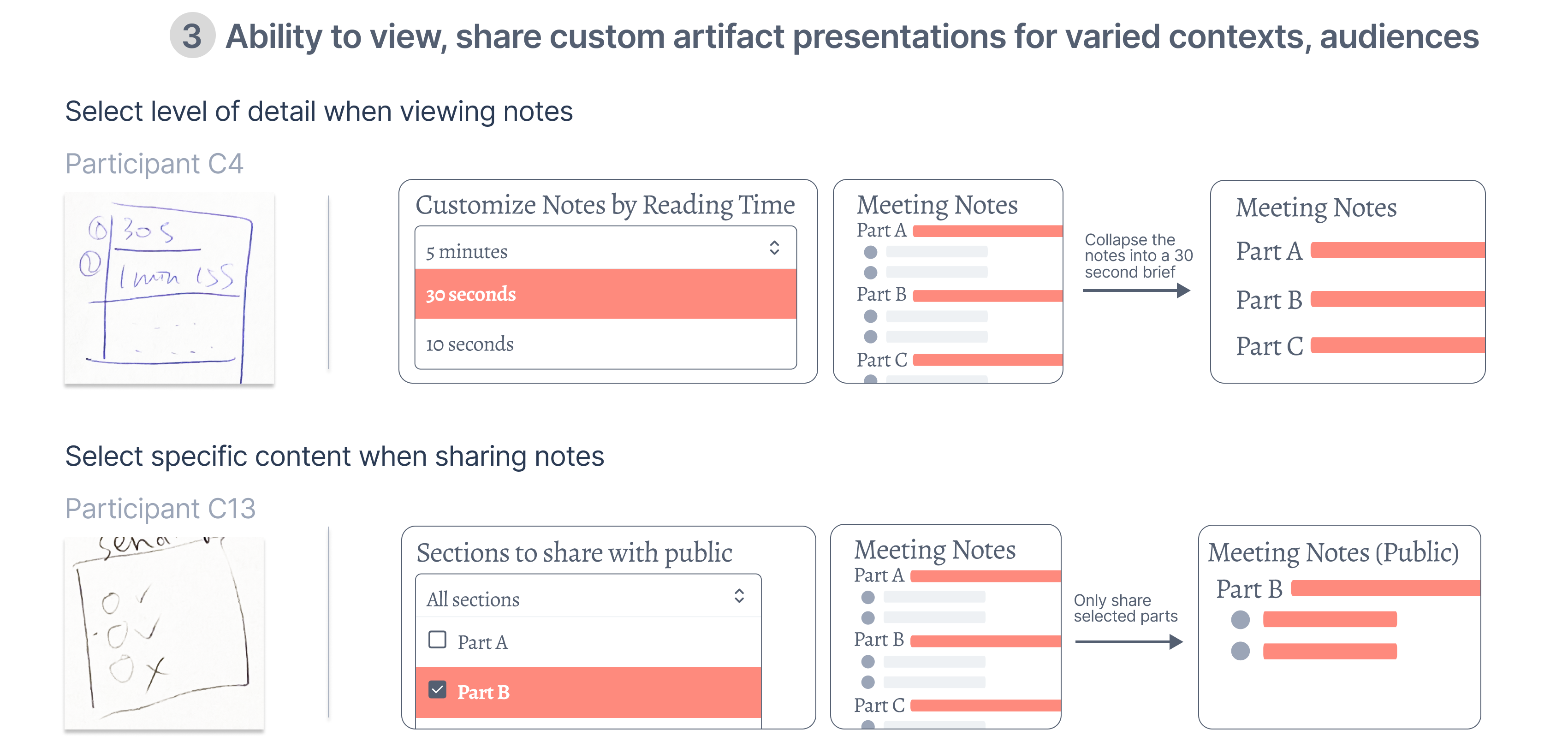}
    \caption{Design Principle \#3: Ability to view, share custom artifact presentations for varied contexts,  and audiences. We show two design examples: select the level of details when viewing notes (top) and select specific content when sharing notes (bottom).}
    \label{fig:DD3}
\endminipage\hfill
\end{figure}

\paragraph{\textbf{\#3: Ability to view, share custom artifact presentations for varied contexts,  audiences.} }
Participants showed strong interest in customizing and personalizing artifacts for different use cases since not all information is necessary for all people, and different team members may play different roles and differ in their information needs. Participants wanted the ability to customize aspects, including the content provided, detail level, and  meeting bridge structure, both as an author when sharing the artifact with others and as a viewer when receiving the artifact.

When sharing meeting bridges with others, some participants wanted to customize the shared content for privacy reasons. For example, C16 wanted to take private notes for himself to explain some terms discussed in the meeting, which he felt was unnecessary for others in the team. 
Participants such as C13 also wanted to customize the presentation of meeting bridges according to the platform on which it is shared: ``\textit{...after the user clicks share, there will be different icons for different platforms. If you share via email, they [the receiver] will only be able to see the text...If you are sharing in Zoom or Slack or other platforms, the receiver will receive the whole [artifact].}''
Participants also sought flexibility in selecting the level of information detail based on the audience's backgrounds or interests. For example, C11 noted  that the artifact needs  minimal detail when shown to meeting attendees but more detail for those who missed the meeting. C16 felt that information could be tailored according to roles and responsibilities.
Finally, participants wanted customization to personalize their experience. C10 preferred to view meeting content in chronological order if they missed the meeting but preferred a topic-based organization for meetings they attended. 

When it came to designs, C3 proposed an expandable, collapsible design: ``\textit{This document with everything collapsed is like a TLDR. This is what we discussed but people can choose to go to whatever point. For example...[another meeting participant]...is new and she may want to see specific things about what is [company name]...So we have that link and people can go to [it]...So now they get to choose like this is when I need more information.}''
Others (C4, C9, C14) proposed a design where users could adjust a slider that sets the artifact's level of detail. For example, C4 suggested a slider design that filters based on the amount of time the viewer wants to spend (Figure~\ref{fig:DD3}).
To achieve structural flexibility, many participants suggested providing multiple indices in meeting bridges for users to choose from. For example, C9 proposed an interactive interface that could switch between topic- and time-based content organizations (Figure~\ref{fig:DD3}).

\begin{figure}
\minipage{1\textwidth}
  \includegraphics[width=\linewidth]{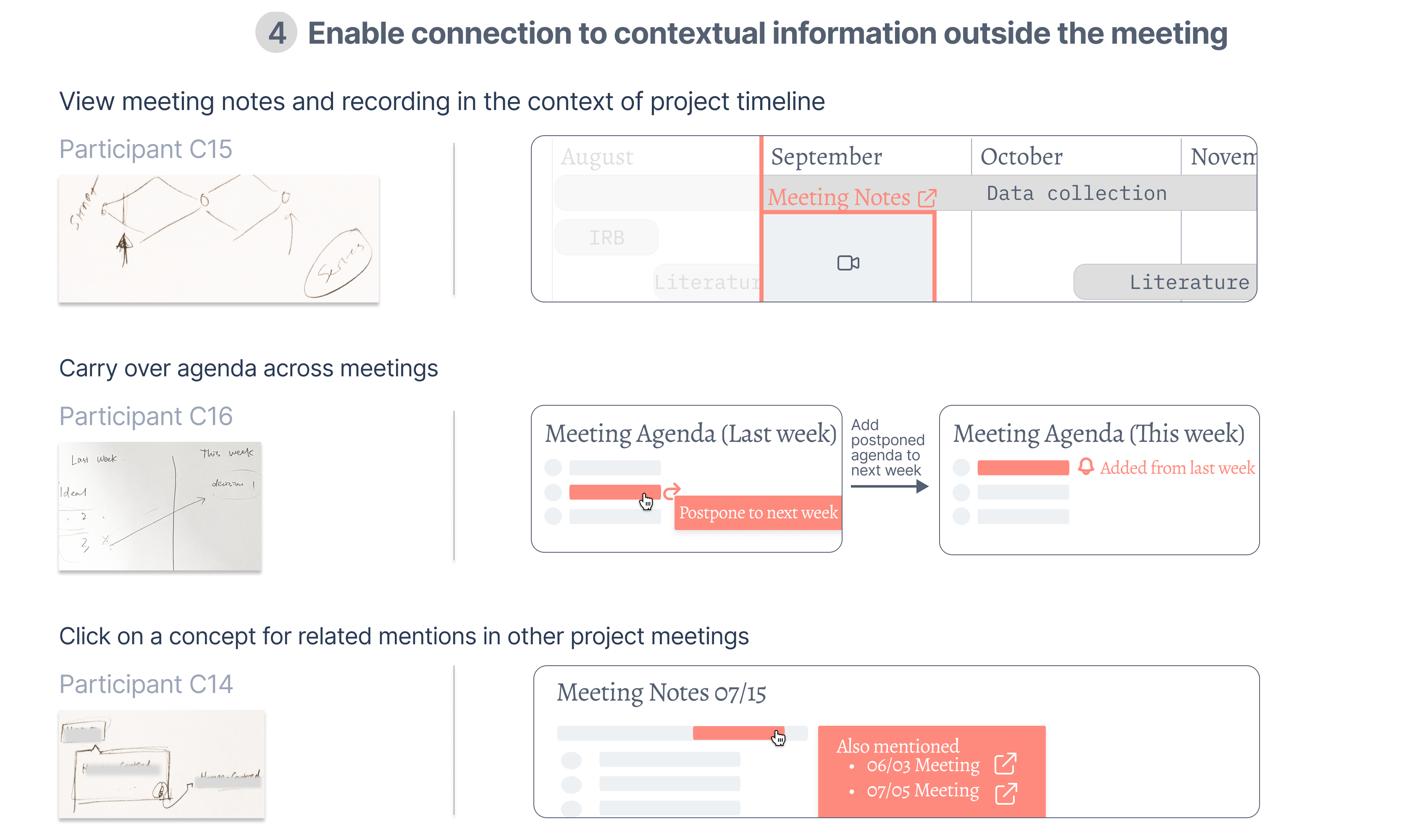}
    \caption{Design Principle \#4: Connection to contextual information outside the meeting. We show three design examples: view meeting materials in the context of the project timeline (top), carry over agenda items across meetings (middle), and click on a concept for related mentions in other project meetings (bottom).}
    \label{fig:DD4}
\endminipage\hfill
\end{figure}

\paragraph{\textbf{\#4: Connection to contextual information outside the meeting.}}
Meetings do not exist in a vacuum; rather, they are often situated in ongoing team collaborations and address issues related to a broader context (e.g., for a larger project or recurring meetings). As a result, participants wanted to contextualize artifact information for  sensemaking and for follow-on collaborations. 

Several participants thought visualizing the relationship between the current and other project meetings would be helpful. C16 proposed providing a project timeline overview that could help the team understand each meeting's goals based on the stages of the current project (Figure \ref{fig:DD4}). C7 and C15 each proposed a design where users could see multiple weeks of meeting bridges at a glance. C15 thought a multi-week design could help him keep track of items raised in previous discussions that were not followed up on (Figure~\ref{fig:DD4}). For C14, when discussion on one topic spans multiple meetings, viewing all meeting bridges together would let them make needed connections. Participants also expressed a desire to situate artifacts in their personal context. C11 proposed a personal multi-meeting dashboard so she could prioritize tasks across different meetings.

Another type of context mentioned by participants is common but implicit knowledge shared by the team, such as team-specific jargon. Here, C16 proposed to link terms and concepts to the previous instances where they were discussed (Figure~\ref{fig:DD4}). He mentioned a case where meeting attendees used the term \textit{persona} in a meeting to describe a specific paper they discussed, not in its usual sense. Similarly, C14 suggested a shared team dictionary to define team-specific terms. 

Finally, participants also wanted to contextualize meeting bridges in their established workflow. Per C13: ``\textit{During the meeting [the artifact] serves as a note-taking sheet, but after the meeting it's a reminder for people who want to go back to the meeting content and follow up.}'' When artifacts are used as reminders of the next steps, participants wanted to integrate them into their calendar or task management system. 

\begin{figure}
\minipage{1\textwidth}
  \includegraphics[width=\linewidth]{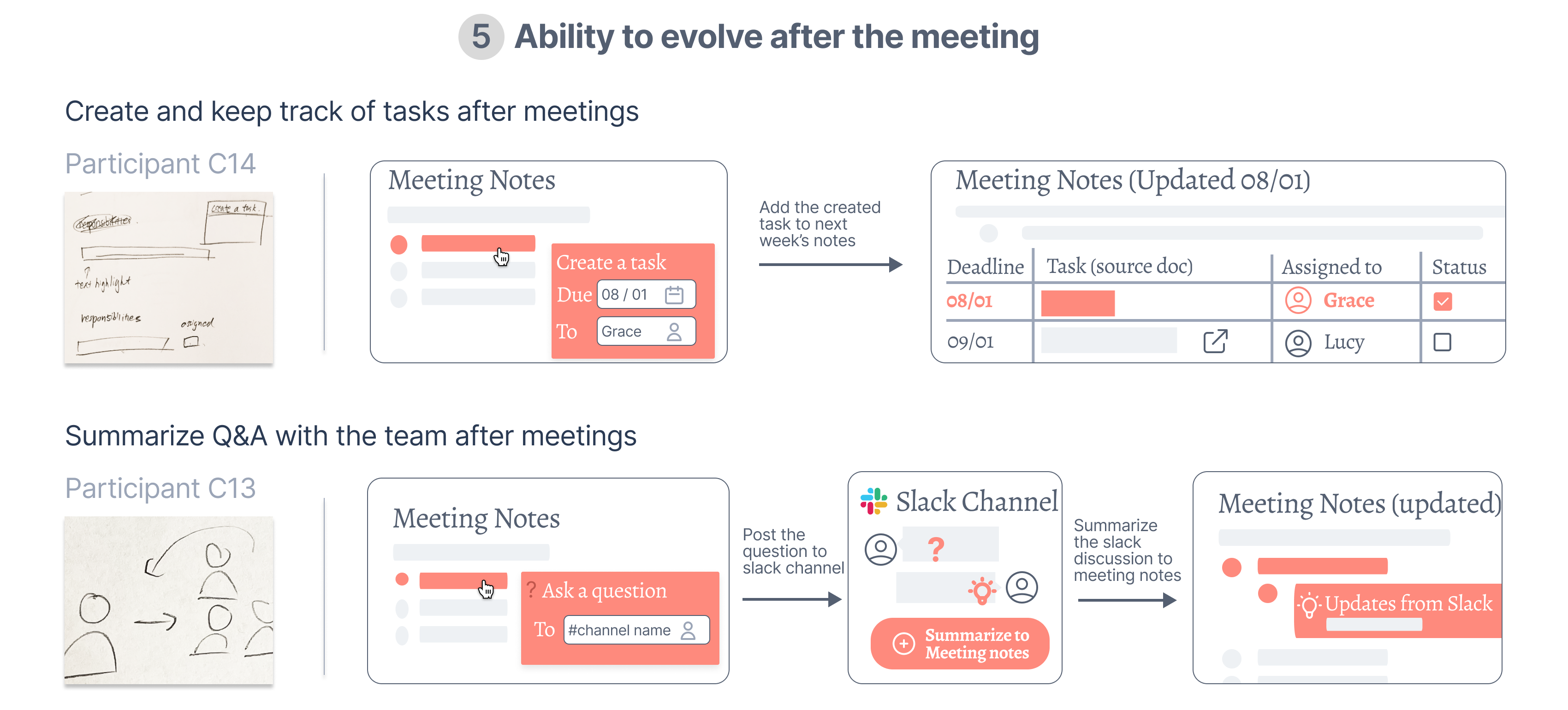}
    \caption{Design Principle \#5: Ability to evolve after the meeting. We show three design examples: create and keep track of tasks after meetings (top) and Q\&A with the team after meetings (bottom).}
    \label{fig:DD5}
\endminipage\hfill
\end{figure}

\paragraph{\textbf{\#5: Ability to evolve after the meeting.}}  \label{sec:DD5}
A team's collaboration typically does not end when the remote synchronous meeting ends. As a result, information captured from a meeting could change after the meeting concludes. For this reason, participants wanted artifacts to be update-able to reflect the latest information, especially when they are used to facilitate further conversations and collaboration among team members. For example, per C9: ``\textit{Sometimes I come back [to the meeting notes] after a week and I found that at some point we already made a decision on something, [but it's not reflected in the meeting notes]. }'' 
C13 wanted status updates in artifacts from a particular meeting; specifically, he wanted the artifact to reflect whether team members reviewed the information captured from the meeting, which helps him set his expectations for the next team meeting.

Participants also suggested that users who did not attend the meeting may  have questions when reviewing the artifact. Many therefore wanted a Q\&A section in the artifact. Several participants also wanted to include a bot to help answer such questions. C9 offered that if the bot could not answer a particular question, it would notify meeting participants to solicit help (Figure~\ref{fig:DD5}).
Beyond questions, meeting attendees could also have new ideas related to the meeting topic after the meeting ends. Some participants wanted to collect these new thoughts and feedback in the artifact as well. For instance, C13 designed their meeting bridge with a separate section for post-meeting thoughts. Collaborators who access the meeting bridge from various platforms (e.g., via email, Slack, or a collaborative doc) could all contribute to the section, which would then be summarized.

\section{Discussion}\label{tension}

We reflect on the concept of meeting bridges by examining their role in supporting remote collaboration, discussing the feasibility of proposed design ideas, and describing tensions designers might experience when creating meeting bridges given current technology. 

\subsection{Designing Meeting Bridges to Support Remote Team Collaboration}

Our findings indicate that bridging the  transition from remote meetings to asynchronous follow-up activities offers a promising starting point for going beyond meeting representations as just ``\textit{a point in time}''~\cite{WorkTrendIndex} towards representing the broader context of an ongoing collaboration involving many touchpoints. 
The challenges we observe in using notes and recordings for asynchronous collaboration echo prior empirical work such as the high cognitive effort of taking notes in meetings~\cite{peverly2007predicts, piolat2005cognitive, yeung1998cognitive}, the difficulty of browsing  videos~\cite{pavelSceneSkimSearchingBrowsing2015, cesarEnhancingSocialSharing2008}, and privacy concerns with sharing meeting records~\cite{cesarEnhancingSocialSharing2008, whittakerAnalysingMeetingRecords2006a}. Many have tried to address these challenges by leveraging in-situ interactions such as annotation~\cite{mu2010towards, zhangMakingSenseGroup2018} or automated methods~\cite{nenkova2011automatic} to add structures to help people make sense of information~\cite{chen2020multi, zhu2020hierarchical}. 
Our work reframes these prior issues and designs to consider them in the context of building \textit{bridges} to support asynchronous collaboration with a concluded synchronous event. By strengthening such bridges, we argue that we can de-emphasize participation in synchronous meetings, alleviating meeting fatigue for team members while avoiding creating more asynchronous work or causing collaboration breakdowns.
With this new frame, we demonstrate how existing systems can embody aspects of our design principles for meeting bridges in Table~\ref{tab:example-table}. 
We notice that many current systems address principles that support users in presenting and accessing rich media information (\#1, \#2), but few systems support users in customizing, sharing, and updating meeting information after the meetings (\#3, \#5), or connecting meeting contents with external information (\#4). 


\begin{table}
\footnotesize
\setlength{\tabcolsep}{2mm}{
\begin{tabular}{p{5cm}p{8cm}}
\toprule
\textbf{Design principles} &
  \textbf{Systems that exemplify the principle} \\ \midrule
\#1: Incorporation of multiple data types and rich media presentations &
   Integrated multi-media platforms such as Classroom 2000~\cite{abowd1999classroom}, Teamspace~\cite{geyer2005towards}; Dashboards such as EMODASH~\cite{ez-zaouiaEmodashDashboardSupporting2020}, Ferret~\cite{cremersProjectBrowserSupporting2007}; Interactive document such as ~\cite{vega-oliverosThisConversationWill2010}\\ \midrule
\#2: Structure(s) to allow users to explore information &
  Auto-generated playback plans such as HyperMeeting~\cite{girgensohnGuidingUsersAsynchronous2016a}; Visual summary such as MeetingVis~\cite{shi2018meetingvis}, Video Manga~\cite{uchihashiVideoMangaGenerating1999};  Interactive video summary such as ElasticPlay~\cite{jinElasticPlayInteractiveVideo2017}; Searchable meeting materials such as Rough `n' Ready~\cite{colbath2000spoken}; User-generated index such as Hotspots~\cite{kalnikaiteMarkupYouTalk2012b} and Filochat~\cite{whittakerFILOCHATHandwrittenNotes1994} \\ \midrule
\#3: Ability to view and share custom presentations of the artifact for different contexts and audience & A system to curate videos to share with heterogeneous participants~\cite{cesarEnhancingSocialSharing2008}; A system to share multimedia screen content that's viewable from any source~\cite{carterWorkCacheSalvagingSiloed2016}
   \\ \midrule
\#4: Ability to enable connection to contextual information outside the meeting &
  Ability to create link across meetings such as HyperMeeting~\cite{girgensohnGuidingUsersAsynchronous2016a} and Time Travel Proxy~\cite{tangTimeTravelProxy2012} \\ \midrule
\#5: Ability to evolve after the meeting &
  Asynchronous video sharing system such as Video Threads~\cite{barksdaleVideoThreadsAsynchronous2012}; Interfaces that collects users' post-meeting feedback for automated meeting understanding~\cite{ehlenMeetingAdjournedOffline2008}\\ \bottomrule
\end{tabular}
}
\caption{Example systems that exemplify each design principle for the design of successful meeting bridges.}
\label{tab:example-table}
\end{table}

To demonstrate how designers can leverage our principles to design more successful bridges, we apply relevant design principle(s) to each of the asynchronous uses for meeting information that we uncovered in our first study, in order to derive concrete ideas (Table~\ref{tab:designGoals}). For example, to design a meeting bridge to onboard new members who join an ongoing project, our design principles suggest designs such as 
 linked multi-media artifacts to help new members to absorb information more effectively, linking team-specific jargon to where they are defined or brought up in prior documents, and a feature for new members to ask follow-up questions based on meeting artifacts. 


\begin{table}
\footnotesize
\begin{tabular}{p{2.9cm}  | p{9.5cm}}
\toprule
\textbf{Asynchronous Uses}  & \textbf{Design Ideas and Relevant Design Principles} \\ \midrule

\begin{tabular}{l}
Personal or group \\archive
\end{tabular}

& As an archive, meeting bridges should support users to access information on-demand, with specific objectives guiding their search. The bridges could incorporate advanced search and indexing functions \textbf{(DP~\#2)} and integrate multiple interconnected data types \textbf{(DP~\#1)} to support user-driven exploration. For example, customizable smart indexes could enable users to filter through multimedia content in ways that resonate with their personal mental model of the meeting. \\  \hline

\begin{tabular}{l}
Personal or group \\task reminder
\end{tabular}

& As task reminders, meeting bridges should be action-oriented and updated with the latest information. It should integrate reminders into the overall project workflow \textbf{(DP~\#4)} and enable users to update task statuses easily \textbf{(DP~\#5)}. A possible design idea is to show a brief task description on calendars or group workspaces (e.g., Slack), with clear deadlines, relevant team members and completion status that are visible to all members.
\\ 
\hline

\begin{tabular}{l}
Group onboarding/\\inclusion
\end{tabular}
 &
To support onboarding, meeting bridges should highlight key takeaways and decisions and allow users to explore details or seek clarifications based on their interests or the perceived importance of information. For example, to help people catch up after missing a meeting, meeting bridges could offer a default overview highlighting main themes using short video snippets of the key moments of the meeting \textbf{(DP~\#1)}, which help users gain a quick grasp of the main ideas. Users could then dive deeper based on their preferences via features like personalized Q\&A \textbf{(DP~\#3)}. To help onboard new members joining the group, meeting bridges could clarify unfamiliar terms by referencing relevant segments from past meetings \textbf{(DP~\#4)}.
\\  \hline

\begin{tabular}{l}
Additional\\Sensemaking
\end{tabular}
 
 &
To support additional sensemaking, meeting bridges should enable users to link captured information across several meetings and with broader contextual details, as well as collaboratively refer to and restructure this information. Drawing on the design principles, a possible design could highlight common themes from recent meetings or related projects \textbf{(DP~\#4)}. It could also offer flexible customization options (e.g., different views) to support users in organizing information in ways that best support their understanding and analysis \textbf{(DP~\#3)}.
 \\
 \hline

\begin{tabular}{l}
Starting point for\\follow-on 
collaboration
\end{tabular}
 
&
To support collaboration beyond the meeting, meeting bridges should provide structure to subsequent synchronous meetings or asynchronous discussions and ensure users' ability to access the information in different contexts. A possible design could be to share video snippets and notes from unresolved discussions to dedicated asynchronous discussion spaces (e.g., Slack) to encourage continuous discussions outside live meetings \textbf{(DP~\#4)}. Updates from these discussions would then be reflected on the meeting bridges to maintain a continuous view of the progress \textbf{(DP~\#5)}.
 \\ \bottomrule
\end{tabular}
\caption{Design idea examples generated by applying design principle(s) to each of the asynchronous uses we uncovered for meeting information.}
\label{tab:designGoals}
\end{table}

\subsection{Technical Considerations in Implementing Meeting Bridges}

Technical challenges could be encountered when building artifacts that meet users' expressed needs. For example, indices that provide structure for existing meeting bridges are often created or extracted manually, adding a burden for users. Similarly, customized views of meeting bridges are often defined manually by users or system developers, which could limit the extent to which the meeting bridge fulfills personal needs. 

The rapidly improving capabilities of emerging technologies, however, introduce new opportunities to mitigate existing challenges. For example, with the growing capabilities of large language models (LLMs) to recognize patterns and generate content based on multimodal data, one can imagine using LLMs to identify latent patterns in meetings and create indices or summaries to help users navigate meeting bridges based on their personal preferences. Further, current multi-media data captured in remote meetings are mostly text or video-based. As sensors of different types become more ubiquitous, one can imagine also including emotion or gaze data that enriches meeting bridges, given appropriate privacy guardrails, of course. Moreover, there could also be potential to combine the strengths of human and machine intelligence to produce more refined meeting bridges with less effort~\cite{jinElasticPlayInteractiveVideo2017,wu2022survey}. 

Additionally, enabling information to flow between collaboration and productivity tools is an essential prerequisite for implementing meeting bridges that bridge from meeting software to other platforms for collaboration. However, closed platforms may create barriers to this. The increasing popularity of open APIs and tools such as Zapier, which provide no-code or automated solutions to connect across platforms, can reduce the cost of manually transferring and transforming information from meeting platforms to other platforms where follow-up activities occur.

\subsection{Limitations and Future Work}

This paper focuses on bridging from synchronous meetings to asynchronous interaction. The other side of the bridge, artifacts for bridging asynchronous communication to  synchronous interactions, should also be considered.  For example,  team participants in a remote synchronous meeting may want to connect to topics discussed in older chat conversations, documents, and other artifacts generated in an asynchronous setting. While this could present some interesting new opportunities to improve live meeting conversations, we chose the synchronous to asynchronous bridge since our formative study demonstrated that this was an area of greater need for participants.

This work also focuses on artifacts that people typically create from remote synchronous video and audio calls. However, people also engage synchronously in collaborative documents (e.g., Google Docs), shared canvases (e.g., Miro, Figma), or virtual spaces (e.g., Gather.town), often while also engaging in synchronous conversations via a video or audio call. Even within video conferencing software such as Zoom, there are additional ways to synchronously collaborate, such as a whiteboard or polls.
Future work could consider how to incorporate information created from these other kinds of shared spaces into a meeting bridge. Some ideas participants expressed in our co-design sessions could nicely pair with these artifacts. For instance, using the idea to create a mind map of topics discussed in a video call, map nodes could directly integrate with the contents of a shared canvas being edited synchronously by members of the call.

A wide variety of meetings for diverse purposes occur in the wild. Although we collected survey data from a sample of participants with diverse backgrounds, our interview data and co-design data were collected from a population that is highly educated and involved in knowledge-intensive work. In addition, we chose to focus our investigation on remote meetings that involve discussion and require follow-up work, and followed this criterion both in the design of the study material and the recruitment of study participants. However, there could be more nuanced behavioral differences when it comes to different meeting types, organizational cultures, attendee roles, etc. Due to the limited scope of our data collection, we do not have data to dive into such nuance and acknowledge that our findings are most suitable to interpret remote meetings that are recurrent in professional contexts with long-term and familiar members, involving brainstorming and discussion. We encourage future work to collect more in-depth empirical evidence to understand the nuance across different meeting setups. Similarly, we acknowledge that our study participants are limited to the U.S. Given that the work culture in different geographical areas differs, it is likely that some of our findings would not generalize to other geographical areas. 

Finally, our co-design study led to the identification of five design principles that can be used to guide design ideas for meeting bridges. However, one should be cautious in using them as a comprehensive checklist, and future research should evaluate these principles quantitatively to understand their effectiveness in facilitating follow-up activities after remote meetings.  

\section{Conclusion}
We explored the current practices, roles, and challenges affecting post-meeting use of information captured from remote meetings. We identified five roles information plays in this regard:  archive, task reminder, support for onboarding and group inclusion, additional sensemaking, and a starting point for follow-on collaboration. We observed that current challenges in capturing, consuming, and sharing information artifacts prevent them from realizing their full potential in supporting these roles. Given the potential uses and challenges with making use of meeting information we encountered via our survey and interviews, we proposed the concept of \textit{meeting bridges}, which encapsulate information artifacts that bridge from remote meetings to asynchronous collaboration. We clarified this concept by identifying five design principles that meeting bridge designers should consider---principles that can guide the future design of information artifacts that help users access and engage with post-meeting information towards reducing meeting and collaboration overload.

\begin{acks}
We would like to thank our study participants and anonymous reviewers. We would also like to thank Loren Terveen, Charlotte Lee, Gary Hsieh, Ye Yuan, and members of the Social Futures Lab for their valuable feedback on the drafts of this paper. 
The work is partly supported by a grant from Microsoft's New Future of Work Initiative.
\end{acks}

\bibliographystyle{ACM-Reference-Format}
\bibliography{reference.bib}

\end{document}